\documentclass[10pt]{article}

\usepackage{mytex}

\newcommand {\Sx}{S_x}
\newcommand {\Sy}{S_y}
\newcommand {\Dx}{D_x}
\newcommand {\Dy}{D_y}

\newcommand {\Sz}{S_{z}}
\newcommand {\Dz}{D_z}

\newcommand {\Dzz}{ D_{zz}}
\usepackage{eepic}
\begin{document}
 
\title{Optimal Boundary Conditions for ORCA-2 Model. }
\author{ Eugene Kazantsev}



\maketitle

\begin{abstract}
A 4D-Var data assimilation technique is applied to a ORCA-2 configuration of the NEMO in order to identify the optimal parametrization of the boundary conditions on the lateral boundaries as well as  on the bottom and on the surface of the ocean.  The influence of the boundary conditions on the solution is analyzed as in the assimilation window and beyond the window. It is shown that  optimal conditions for  vertical operators allows to get stronger and finer  jet streams (Gulf Stream, Kuroshio) in  the solution. Analyzing  the reasons of the jets reinforcement, we see   that the major impact of the data assimilation is made on the parametrization of the bottom boundary conditions for lateral velocities $u$ and $v$. 

Automatic generation of the tangent and adjoint codes  is also discussed. Tapenade software  is shown to be able to produce the adjoint code that can be used after a memory usage optimization. 

\end{abstract}

 


\section{Introduction}

Variational data assimilation technique,  first proposed in \cite{Ledimet82}, \cite{ldt86}, is based on the optimal control methods \cite{Lions68} and perturbations theory   \cite{Marchuk75}. This technique allows us to retrieve an optimal data for a given model from heterogeneous observation fields. Since the early 1990's, many mathematical and geophysical teams are involved in the development of the data assimilation strategy. One can cite many papers devoted to this problem, as  in the domain of development of different methods for the  data assimilation  and also in the domain of its applications to the atmosphere and oceans.  

In the beginning, data assimilation methods were intended to identify and reconstruct an optimal initial state of the model. However, the idea that other model's parameters should  also be identified by data assimilation has also been studied and discussed in numerous papers.  
One can cite several examples of using  data assimilation  to identify the bottom topography of simple models (\cite{LoschWunsch}, \cite{Heemink}, \cite{assimtopo}), to control open boundary conditions in coastal and regional models (\cite{shulman97}, \cite{shulman98}, \cite{Taillandier}, \cite{Brummelhuis}), boundary conditions on rigid boundaries (\cite{fxld-mo}, \cite{Leredde}  \cite{Lellouche}, \cite{assimbc1}, \cite{sw-lin}, \cite{sw-nl})  and to determine  other parameters of a model (\cite{zou}, \cite{panchang}, \cite{chertok}). 

It was pointed out in  \cite{navon97}  that  the problem of adjoint parameters identification is frequently ill-posed. This fact remains valid when boundary conditions are considered as a control parameter. It was shown in \cite{assimbc1} that the presence of non-null kernel of the Hessian  results in a non-unique  choice of optimal boundary conditions. However, all sets in the kernel from the kernel are equivalent: they provide the same (or almost the same) cost function's value and almost the same evolution of the model's solution after the end of assimilation. Consequently, we can not  talk about physical meaning of the optimal conditions. Indeed, any value from the kernel can be obtained as the result of the minimization procedure while only one of them may have a physical meaning. In this paper we take this ambiguity into account and consider the adjoint boundary condition estimation as a simple compensation of the model errors rather than real identification of model parameters. 

Particular attention is payed to the design of the tangent and adjoint codes. As it has been shown in  \cite{sw-lin}, the derivative of the model with respect to boundary conditions is two or three times longer (as well as in terms of the development, the number lines of the code and the necessary CPU time) than the derivative used to control the initial point of the model. The usual tangent model describes a linear evolution of a small perturbation to a model solution while the derivative with respect to other parameter   includes also a block that introduce the perturbation into the model. This block may be at least as long and complex   as the whole derivative with respect to initial conditions.

That's why we focus our attention on the automated differentiators. In this paper tangent and adjoint   codes have been obtained by Tapenade software \cite{Hascoet04}. Current version of this software reveals to be able to produce the derivative of such a complex code as ORCA-2.  

The paper is organized as follows. In the second section, $2^0$ global ocean configuration (ORCA-2)\footnote{http://www.mercator-ocean.fr/eng/science/composantes-systemes/modelisation/orca2}   of the Nucleus for European Modelling of the Ocean (NEMO)\footnote{http://www.nemo-ocean.eu/} model   is described.  The third section is devoted to the data assimilation technique and to  the Automatic Differentiation Engine  Tapenade\footnote{http://www-tapenade.inria.fr:8080/tapenade/index.jsp} software to generate the adjoint and to the optimization of the generated code. Results and discussion are presented in the fourth section. 

\section{ORCA-2 configuration of the NEMO and its discretization.}

ORCA-2 configuration of the Nucleus for European Modelling of the Ocean \cite{madec-nemo} is used in this paper. This is $2^0$ global ocean configuration based on the OPA 8.2  \cite{madec-opa} primitive equations model. 

 Its domain extends from $78^0$S to $90^0$N.   The model grid counts $182\tm 149 \tm 31$ nodes. Vertical discretization is performed on  $z$ levels with the partial step approximation of the bottom cell \cite{partsteps}. Vertical mixing is achieved using the turbulent kinetic energy (TKE) scheme described in \cite{blanke}.

Spatially discretized equations of the ORCA-2 model are written as follows:

\beqr
\der{u}{t}&=&
 \underbrace{
  \biggl( \Sx\Sy v\biggr)\Sy(\omega+f)- \Dx\fr{\Sx u^2+ \Sy v^2}{2}-\Sz\biggl(\Sx w \Dz u\biggr)
 }_{\mbox{ Advection}}+
 \underbrace{
   { D_x}{A^{xy}_u\xi}+{ D_y}{A^{xy}_u\omega}
 }_{\mbox{ Lateral diffusion}}   +
\nonumber\\ &+& 
 \underbrace{
   g\int_{surface}^z\Dx \Sz \rho(x,y,\zeta)d\zeta
 }_{\mbox{Hydrostatic  Pressure grad.}}  +   
 \underbrace{  
   \Dzz A^z_u u
 }_{\mbox{ Vertical diffusion}}   + 
 \underbrace{ 
 g \Dx(\eta+T_c\partial_t\eta)
  }_{\mbox{ Surface Pres. Grad.}}   
 \label{1.1}\\ 
 \der{v}{t}&=&
 \underbrace{
  \biggl( \Sx\Sy u\biggr)\Sx(\omega+f)- \Dy\fr{\Sx u^2+ \Sy v^2}{2}-\Sz\biggl(\Sx w \Dz v\biggr)
 }_{\mbox{ Advection}}+
 \underbrace{
   { D_x}{A^{xy}_v\xi}+{ D_y}{A^{xy}_v\omega}
 }_{\mbox{ Lateral diffusion}}   +
\nonumber\\ &+& 
 \underbrace{
   g\int_{surface}^z\Dy \Sz \rho(x,y,\zeta)d\zeta
 }_{\mbox{Hydrostatic  Pressure grad.}}  +   
 \underbrace{  
   \Dzz A^z_v v
 }_{\mbox{ Vertical diffusion}}   + 
 \underbrace{ 
 g \Dy(\eta+T_c\partial_t\eta)
  }_{\mbox{ Surface Pres. Grad.}}   
 \label{1.2}\\ 
 \der{T}{t}&=&
  \underbrace{
   -\Dx (u\Sx T)-\Dy (v \Sy T)-\Dz (w \Sz T)
     }_{\mbox{Advection}}   +
 \underbrace{ 
   A^{xy}_T\biggl(D_x D_xT+D_yD_yT\biggr)
  }_{\mbox{Lateral diffusion}}     +
\nonumber\\
 &+& 
  \underbrace{
    \Dzz A^z_T T
  }_{\mbox{Vertical diffusion}}       
    +\mbox{\scriptsize Solar Radiation}+\mbox{\scriptsize Geothermal Heating}+\mbox{\scriptsize BBL} \label{1.3} \\
 \der{s}{t}&=&
  \underbrace{
   -\Dx (u\Sx s)-\Dy (v \Sy s)-\Dz (w \Sz s)
     }_{\mbox{Advection}}   +
 \underbrace{ 
   A^{xy}_T\biggl(D_x D_x s+D_yD_y s\biggr)
  }_{\mbox{Lateral diffusion}}     +
\nonumber\\
 &+& 
  \underbrace{
    \Dzz A^z_T s
  }_{\mbox{ Vertical diffusion}}       
   +\mbox{\scriptsize BBL} \label{1.4} \\
  \der{\eta}{t}&=&   w(x,y,z=\mbox{surface}) \label{1.5} \\
 \xi&=& {\Dx}u+{\Dy}v,\hspace{3mm}\omega={\Dy}u-{\Dx}v,\hspace{3mm}
 w=\int_{bottom}^z \xi(x,y,\zeta) d\zeta,\hspace{3mm}
 \rho=\rho(T,s)\label{1.6}
 \eeqr
 where operators $D$ and $S$ are derivatives and interpolations on the Arakawa C-grid. These operators will be discussed below in details. 
 
 The set of variables in this system is composed of  $u,v$ and $w$ that represent  zonal, meridional and vertical velocity components, $T$ and $s$ denote the potential temperature and salinity,   $\xi$ and $\omega$ are horizontal divergence and vorticity, $\eta$ is the sea surface elevation and $\rho$ is the density anomaly that is defined as a function of the temperature and salinity by  the state equation.  As one can see, $u,v,T,s,\eta$ are prognostic variables while  $w, \xi, \omega$ and $\rho$ are diagnostic ones. 
 
 Among parameters in these equations, we can see $f=2\Omega\sin(latitude)$ that represents  the Coriolis parameter, gravity acceleration $g=9.81\fr{m}{s^2}$, lateral diffusion coefficients  
 $$A^{xy}_T=2000\fr{m^2}{s},\;\;A^{xy}_u=\left\{\begin{array}{l}40000\fr{m^2}{s}\mbox{latitude}>15^0\\2000\fr{m^2}{s}\mbox{latitude}<15^0, \end{array}\right. . $$
 
 To calculate coefficients of the vertical diffusion $A^z_u, A^z_v, A^z_T, A^z_s$ we use the turbulent closure scheme accompanied by the  double diffusive mixing and enhanced vertical diffusion approximations. 
 
The  turbulent closure scheme is applied to solve the problem of statically unstable density profiles. 
 The vertical eddy viscosity and diffusivity coefficients are computed from a TKE turbulent
closure model based on a prognostic equation for $\bar{e}$, the turbulent kinetic energy  \rf{tke},
and a closure assumption for the turbulence length scales \cite{madec-opa}. 
 \beqr
 \der{\bar{e}}{t}&=&
A^z_{v}\left[\left(\der{u}{z}\right)^2+\left(\der{v}{z} \right)^2 \right]
-A^z_{T}\,N^2+\der{}{z}\left[A^{z}_u\der{\bar{e}}{z} \right]
- c_\epsilon \;\frac{\bar {e}^{3/2}}{l_\epsilon }
  \label{tke} \\
         A^z_{u}=A^z_{v} &=& \max(A^z_0,C_k\  l_k\  \sqrt {\bar{e}})\label{tke1}  \\
         A^z_{T} &=& A^z_{u} / P_{rt} \label{az}
\eeqr
where $N$ is the local Brunt-Vais\"{a}l\"{a} frequency which is calculated as a function of $T$ and $s$. Parameters 
$l_{\epsilon }$ and $l_{\kappa }$ are the dissipation and mixing length scales, 
$P_{rt} $ is the Prandtl number. The constants $C_k = \sqrt {2} /2$ and 
$c_\epsilon = 0.1$ are designed to deal with vertical mixing at any depth 
\cite{Gaspar1990}.  Following \cite{blanke}, $P_{rt} $  
is defined as a function of the local Richardson number, $R_i $:
$$
P_{rt} = \left\{ \begin{array}{ll}
                    1 &      \mbox{if }\ R_i \leq 0.2 	\\
                    5\,R_i &      \mbox{if }\ 0.2 \leq R_i \leq 2 	\\
                    10 &      \mbox{if } 2 \leq R_i 
            \end{array}\right.
$$

In frames of the enhanced vertical diffusion parameterization, we assign very big values to the vertical eddy mixing coefficients  in regions where the stratification is unstable (i.e. when the Brunt-Vais\"al\"a frequency
is negative) \cite{Lazar99}. This is done  on both momentum $u,v$ and tracers $T,s$:
\beq
A^z= \{\mbox{ if } N^2<0 \mbox{ then } A^z=100  \} \label{az-enhvisc}
\eeq

Double diffusion occurs when relatively warm, salty water overlies cooler, fresher
water, or vice versa. They contribute to diapycnal mixing in extensive regions
of the ocean. The  parameterization of such phenomena was included 
in a global ocean model and it was shown that it leads to relatively minor changes in circulation
but exerts significant regional influences on temperature and salinity   in \cite{Merryfield99}.

\beqr
A^z_{T} &=& A^z_{T}+ 
\left\{\begin{array}{ll}
0.7 \frac{10^{-4}}{R_\rho (1+(R_\rho / 1.6)^6)}&\mbox{if  $R_\rho > 1$  }  \\
 A_{ddm}&\mbox{if  $0<R_\rho < 1$  } \\
0&\mbox{otherwise} 
\end{array}\right. 	
\nonumber\\
A^z_{s} &=& A^z_{s}+
\left\{\begin{array}{ll}
 \frac{10^{-4}}{1+(R_\rho / 1.6)^6}&\mbox{if  $R_\rho > 1$  }  \\
 \left( 1.85\,R_{\rho} - 0.85 \right)A_{ddm}&\mbox{if  $0.5 \leq R_\rho<1$  } \\
  0.15 R_\rho A_{ddm}& \mbox{if  $\ \ 0 < R_\rho<0.5$ } \\
\end{array}\right. 	
\label{az-ddm}
\\
A^z_{u}&=&\max(A^z_{u},A^z_{T},A^z_{s})\nonumber\\
A^z_{v}&=&\max(A^z_{v},A^z_{T},A^z_{s})\nonumber
\eeqr
where $A_{ddm}=1.3636\tm 10^{-6} e^{ 4.6 e^{-0.54 (1/R_\rho -1)}}$ and
  $R_\rho$ is the buoyancy ratio $R_\rho \sim \partial_z T /  \partial_z S$.

The term  $T_c\partial_t\eta$ in the equations \rf{1.1} and \rf{1.2} is introduced to dump the external gravity waves. These waves are fast so their timescale is short with respect to other processes described by the primitive equations. Explicit resolution of these waves requires too short time step which is unnecessary to all other physics. Consequently,   the filter of temporally unresolved external gravity waves, proposed in \cite{RoulletMadec}, is introduced into the model. The cutoff time  $T_c$ is equal to  one time step of the model.  

The model is discretized on the grid, that is the generalization to three dimensions of the well-known ``C'' grid in 
Arakawa's classification \cite{Mesinger_Arakawa_Bk76}. The arrangement of variables is the same in all directions. 
It consists of cells centered on scalar points ($T$, $s$, $\eta$, $\rho$) with vector 
points $(u, v, w)$ defined in the center of each face of the cells. The relative and 
planetary vorticity, $\omega$ and $f$, are defined in the center of each vertical edge.

 To perform interpolations on this grid and to calculate the derivatives, we introduce operators $S$ and $D$ in the discretized equations \rf{1.1}--\rf{1.6}. These operators plays the key role in this study because they depend on the boundary conditions and introduce them into the model. Despite these operators are usually written as overbars and $\delta$ brackets (see \cite{madec-nemo}, for example), we note them as letters with index in order to emphasize that these are operators under control in this paper. Interpolations are calculated as a  weighted mean of two function values in the adjacent nodes. Weights are defined to be proportional to the grid steps of corresponding cells in order to ensure the second order interpolation of a grid function. These weights are described in details in the  \cite{madec-nemo}, but we omit them in this paper to concentrate our attention on the principal features. Thus, writing operators in a simplified way, we assume both the argument and the result of the interpolation operator and the derivative are  multiplied by an appropriate weight.

At each grid-node in the ocean they  we write interpolations $\Sx, \Sy$ and $\Sz$ in a  common way: 
\beqr
(\Sx u)_{i+1/2,j,k}=\fr{u_{i,j,k}+u_{i+1,j,k}}{2},\; 
(\Sy u)_{i,j+1/2,k}=\fr{u_{i,j,k}+u_{i,j+1,k}}{2}, \; \nonumber \\
(\Sz u)_{i,j,k+1/2}=\fr{u_{i,j,k+1}+u_{i,j,k}}{2}
\label{s-mid}
\eeqr
Following \cite{assimbc1}, \cite{sw-nl}, these expressions are modified in the grid-nodes adjacent to the boundary, i.e. near the continents for interpolations in $x$ and $y$ directions and near the bottom and the surface for vertical interpolation. 

Let us suppose the index $i=0$ corresponds to left rigid  boundary. That means the index $i=1$ is the first grid node in the ocean. In this case, we use the formula
\beq
(\Sx u)_{1/2,j,k}={ \alpha^{Sux^l}_0}+{ \alpha^{Sux^l}_1} u_{0,j,k}+{\alpha^{Sux^l}_2} u_{1,j,k}
\label{s-1}
\eeq
to calculate the interpolated value of $u$ at the point $1/2,j,k$, where scalar variables $T,s,\rho$ are defined. 
The value of $(\Sx u)_{N-1/2,j,k}$ near the right boundary is calculated by the similar formula, but with different coefficients $\alpha^{Sux^r}_0, \alpha^{Sux^r}_1, \alpha^{Sux^r}_2$. 
 
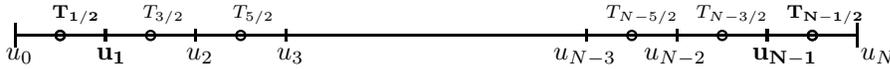
\begin{figure}[h]
\setlength{\unitlength}{0.8mm}
\newcount\indi
\newcount\num
\begin{picture}(150,20)
\linethickness{0.25mm}
\put(5,10){\line(1,0){140}}
\put(5,8){\line(0,1){4}}
\put(145,8){\line(0,1){4}}
\thicklines

\put(5,9){\line(0,1){2}}
\multiput(35,9)(15,0){2}{\line(0,1){2}}
\put(2,6){  $u_{0}$ }
\indi=1
\multiput(32,6)(15,0){2}{ 
\global\advance\indi by 1 $u_{\the\indi}$ }

\put(130,9){\line(0,1){2}}
\multiput(115,9)(-15,0){2}{\line(0,1){2}}
\put(144,6){ $u_{N}$ }
\indi=1
\multiput(108,6)(-15,0){2}{ 
\global\advance\indi by 1 $u_{N-\the\indi}$ }

\multiput(12.5,10)(15,0){3}{\circle{1.5}}
\indi=0
\scriptsize
\multiput(25,13)(15,0){2}{ 
\global\advance\indi by 1 \num=\indi \multiply\num by 2 \global\advance\num by 1 $T_{\the\num/2}$ }
\multiput(137.5,10)(-15,0){3}{\circle{1.5}}

\indi=0
\multiput(117,13)(-15,0){2}{ 
\global\advance\indi by 1 \num=\indi \multiply\num by 2 \global\advance\num by 1 $T_{N-\the\num/2}$ }
\scriptsize
\put(12.5,10){\circle{1.5}}
\put(137.5,10){\circle{1.5}}
\put(10,13){  $\mathbf{T_{1/2}}$ }
\put(132,13){  $\mathbf{T_{N-1/2}}$ }
\normalsize
\put(20,9){\line(0,1){2}}
\put(17,6){  $\mathbf{u_{1}}$ }
\put(130,9){\line(0,1){2}}
\put(126,6){ $\mathbf{u_{N-1}}$ }
\end{picture} 
\caption{ Structure of the horizontal grid. }
\label{xygrid}
\end{figure}

Similarly,   interpolated value of $T,s,\rho$ at points $i=1, i=N-1$  are calculated by 
\beqr
(\Sx T)_{1,j,k}&=&{ \alpha^{STx^l}_0}+{ \alpha^{STx^l}_1} T_{1/2,j,k}+{\alpha^{STx^l}_2} T_{3/2,j,k}
\nonumber\\
(\Sx T)_{N-1,j,k}&=&{ \alpha^{STx^r}_0}+{ \alpha^{STx^r}_1} T_{N-1/2,j,k}+{\alpha^{STx^r}_2} T_{N-3/2,j,k}
\nonumber
\eeqr

Coefficients $\alpha$ play the role of the control variables in this paper. Operators  are allowed to change their properties near boundaries in order to find the best fit with requirements of the model and data.   To assign all control variables   we shall perform data assimilation procedure and find their optimal values.

The first coefficient, $\alpha_0$, is added to the interpolation formula to simulate non-uniform boundary conditions. Coefficient $\alpha_1$ controls the contribution of the physical boundary value that is prescribed to $u_{0,j,k}$. It should be noted here, that if impermeability condition is imposed, $u_{0,j,k}=0$, the interpolation \rf{s-1} is controlled by two $\alpha$ only. 

Similar modification of the interpolation formulas are performed  near the north and the south boundaries in the $y$ direction. Exception is made for the periodical conditions. They are  applied  on the  $78^0E$ longitude and near the North Pole. In this case no modification is made and no control is applied. 

In the vertical direction we control the discretization at two  points: one point near the surface and another one  near the bottom (see \rfg{zgrid}). To calculate the interpolated values of the temperature, for example, we use the following formula:
\beqr
(\Sz T)_{i,j,1}&=&{ \alpha^{SzT^s}_0}+{ \alpha^{SzT^s}_1} T_{i,j,1/2}+{\alpha^{SzT^s}_2} T_{i,j,3/2}
\nonumber\\
(\Sz T)_{i,j,K-1}&=&{ \alpha^{SzT^b}_0}+{ \alpha^{SzT^b}_1} T_{i,j,K-1/2}+{\alpha^{SzT^b}_2} T_{i,j,K-3/2}
\label{sz-1}
\eeqr

\begin{figure}[h]
\setlength{\unitlength}{0.8mm}
\newcount\indi
\newcount\num

\newcount\postrait
\newcount\postxt
\newcount\oldtrait

\linethickness{0.3mm}
\begin{picture}(150,20)

\put(5,10){\line(1,0){55}}
\put(63,10){$\cdots$}
\put(70,10){\line(1,0){75}}

\put(5,8){\line(0,1){4}}
\put(145,8){\line(0,1){4}}
\thicklines

\put(2,6){ $w_{0}$ }

\postrait=5
\put(\the\postrait,16){$\overbrace{\linethickness{0mm}\line(1,0){8}\linethickness{0.3mm}}^{hz_{1/2}}$}
\global\advance\postrait by 8
\postxt=\postrait
\global\advance\postxt by -3
\put(\the\postrait,9){\line(0,1){2}}
\put(\the\postxt,6){ $\mathbf{w_{1}}$ }

\put(\the\postrait,16){$\overbrace{\linethickness{0mm}\line(1,0){12}\linethickness{0.3mm}}^{hz_{3/2}}$}
\global\advance\postrait by 12
\postxt=\postrait
\global\advance\postxt by -3
\put(\the\postrait,9){\line(0,1){2}}
\put(\the\postxt,6){ \bf $w_{2}$ }

\put(\the\postrait,16){$\overbrace{\linethickness{0mm}\line(1,0){16}\linethickness{0.3mm}}^{hz_{5/2}}$}
\global\advance\postrait by 16
\postxt=\postrait
\global\advance\postxt by -3
\put(\the\postrait,9){\line(0,1){2}}
\put(\the\postxt,6){ $w_{3}$ }

\postrait=145
\put(142,6){ $w_{K}$ }
\global\advance\postrait by -25
\put(\the\postrait,16){$\overbrace{\linethickness{0mm}\line(1,0){25}\linethickness{0.3mm}}^{hz_{K-1/2}}$}
\postxt=\postrait
\global\advance\postxt by -3
\put(\the\postrait,9){\line(0,1){2}}
\put(\the\postxt,6){ $\mathbf{w_{K-1}}$ }

\global\advance\postrait by -22
\put(\the\postrait,16){$\overbrace{\linethickness{0mm}\line(1,0){22}\linethickness{0.3mm}}^{hz_{K-3/2}}$}
\postxt=\postrait
\global\advance\postxt by -3
\put(\the\postrait,9){\line(0,1){2}}
\put(\the\postxt,6){ $w_{K-2}$ }
\global\advance\postrait by -19
\postxt=\postrait
\global\advance\postxt by -3
\put(\the\postrait,9){\line(0,1){2}}
\put(\the\postxt,6){ $w_{K-3}$ }


\scriptsize

\postrait=5
\global\advance\postrait by 4
\postxt=\postrait
\global\advance\postxt by -3
\put(\the\postrait,10){\circle{1.5}}
\put(\the\postxt,13){ $\mathbf{T_{1/2}}$ }

\put(\the\postrait,5){$\underbrace{\linethickness{0mm}\line(1,0){10}\linethickness{0.3mm}}_{hz_{1}}$}
\global\advance\postrait by 10
\postxt=\postrait
\global\advance\postxt by -3
\put(\the\postrait,10){\circle{1.5}}
\put(\the\postxt,13){ $T_{3/2}$ }

\put(\the\postrait,5){$\underbrace{\linethickness{0mm}\line(1,0){14}\linethickness{0.3mm}}_{hz_{2}}$}
\global\advance\postrait by 14
\postxt=\postrait
\global\advance\postxt by -3
\put(\the\postrait,10){\circle{1.5}}
\put(\the\postxt,13){ $T_{5/2}$ }

\put(\the\postrait,5){$\underbrace{\linethickness{0mm}\line(1,0){18}\linethickness{0.3mm}}_{hz_{3}}$}
\global\advance\postrait by 18
\postxt=\postrait
\global\advance\postxt by -3
\put(\the\postrait,10){\circle{1.5}}
\put(\the\postxt,13){ $T_{7/2}$ }

\postrait=145
\global\advance\postrait by -13
\postxt=\postrait
\global\advance\postxt by -3
\put(\the\postrait,10){\circle{1.5}}
\put(\the\postxt,13){ $\mathbf{T_{K-1/2}}$ }

\global\advance\postrait by -23
\put(\the\postrait,5){$\underbrace{\linethickness{0mm}\line(1,0){23}\linethickness{0.3mm}}_{hz_{K-1}}$}
\postxt=\postrait
\global\advance\postxt by -3
\put(\the\postrait,10){\circle{1.5}}
\put(\the\postxt,13){ $T_{K-3/2}$ }
\global\advance\postrait by -20
\put(\the\postrait,5){$\underbrace{\linethickness{0mm}\line(1,0){20}\linethickness{0.3mm}}_{hz_{K-2}}$}
\postxt=\postrait
\global\advance\postxt by -3
\put(\the\postrait,10){\circle{1.5}}
\put(\the\postxt,13){ $T_{K-5/2}$ }

\normalsize
\linethickness{0.3mm}

\end{picture} 
\caption{ Structure of the vertical grid. The number of layers depends on the longitude and latitude: $K=K_{i,j}$}
\label{zgrid}
\end{figure}
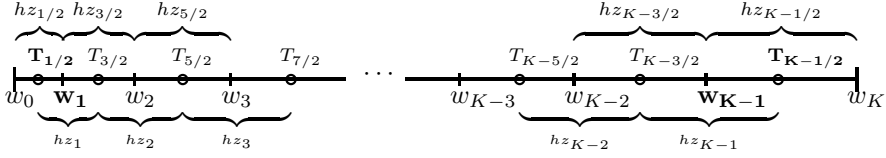

We note here, that the value of the temperature (and other variables defined at the temperature levels) is not extrapolated to nodes $w_0$ and $w_{K}$ where physical boundary conditions must be prescribed. These conditions participate in the interpolation from $w$ levels to $T$ levels 
\beqr
(\Sz w)_{i,j,1/2}&=&{ \alpha^{Szw^s}_0}+{ \alpha^{Szw^s}_1} w_{i,j,0}+{\alpha^{Szw^s}_2} w_{i,j,1}
\nonumber\\
(\Sz w)_{i,j,K-1/2}&=&{ \alpha^{Szw^b}_0}+{ \alpha^{Szw^b}_1} w_{i,j,K}+{\alpha^{Szw^b}_2} w_{i,j,K-1}
\label{sz-2}
\eeqr

Discretized derivatives near the boundary are defined in a similar way as for the horizontal derivatives and for the vertical ones.
\beqr
(\Dx u)_{1/2,j,k}&=&{ \alpha^{Dux^l}_0}-{ \alpha^{Dxu^l}_1} u_{0,j,k}+{\alpha^{Dxu^l}_2} u_{1,j,k}\nonumber\\
(\Dz w)_{i,j,K-1/2}&=&{ \alpha^{Dzw^b}_0}+{ \alpha^{Dzw^b}_1} w_{i,j,K}-{\alpha^{Dzw^b}_2} w_{i,j,K-1}
\label{d-1}
\eeqr
As well as above, impermeability conditions are  prescribed at  lateral boundary  for $u$ and $v$. That means coefficients $\alpha_1$, being multiplied by the vanishing value of the function on the boundary,  are not under control in the experiment. On the other hand, boundary condition on the surface of the ocean,  prescribed for vertical derivatives, are really used in the model and controlled by corresponding coefficient $\alpha^{Dz}_1$. 

Taking into account the fact that at different points optimal boundary conditions may be different, we accept the coefficients $\alpha$ may vary from point to point. Thus, $\alpha^{Dz}$ and  $\alpha^{Sz}$ are considered as functions of the longitude and the latitude. 
Coefficients used in horizontal operators are also allowed to vary from one boundary point to another. 

Along with the derivatives and interpolations, boundary conditions influence also  the second derivatives in the vertical diffusion and reconstruction of the vertical velocity $w$ from the horizontal divergence \rf{1.6}. Discretized equivalent of the \rf{1.6} writes
\beqr
w_{i,j,K-1}&=& \alpha^{w^b}_0 -\alpha^{w^b}_1 hz_{i,j,K-1}\xi_{i,j,K-1}  \nonumber\\
w_{i,j,k-1}&=& w_{i,j,k} -hz_{i,j,k-1}\xi_{i,j,k-1}\quad \forall k: 2\leq k \leq K-1  \label{w}\\
w_{i,j,1}&=& w_{i,j,2}+ \alpha^{w^s}_0 -\alpha^{w^s}_1 hz_{i,j,1}\xi_{i,j,1} 
 \nonumber
 \eeqr

One can see the impermeability condition on the bottom may be violated by the term $\alpha^{w^b}_0$ in the approximation of the vertical velocity near the bottom $w_{i,j,K-1}$ that may allow non-zero vertical velocity.  Coefficient $\alpha^{w}_1$ allows to control the thickness  of the surface and the bottom layers.   

Control of the boundary conditions for the second derivatives $D_{zz}$ in \rf{1.1} is performed in a slightly different  way. We add $\alpha_0$ to the prescribed boundary conditions on the surface and on the bottom
\beqr
\der{u}{z}\bigg\vert_{surface}&=&\alpha_0^{DzzU^s}+\fr{\tau_x}{hz_1\rho_0},\quad
\der{v}{z}\bigg\vert_{surface}=\alpha_0^{DzzU^s}+\fr{\tau_y}{hz_1\rho_0},\nonumber \\
\der{T}{z}\bigg\vert_{surface}&=&\der{S}{z}\bigg\vert_{surface}=\alpha_0^{DzzT^s}\nonumber \\
u\vert_{bottom}&=&v\vert_{bottom}=\alpha_0^{DzzU^b}\quad T\vert_{bottom}=S\vert_{bottom}=\alpha_0^{DzzT^b}
\eeqr
and modify the approximation of the second derivatives at the nodes, adjacent to the surface and to the bottom multiplying  the finite differencing coefficients  by $\alpha_1$ and $\alpha_2$. Thus,for example, approximation of $\dder{u}{z}$ writes

\beqr
\biggl(D_{zz} u\biggr)_{i,j,1/2}&=& \fr{(A^z_u)_{1}}{hz_{1}hz_{1/2}}(\alpha^{DzzU^s}_2 u_{3/2}-\alpha^{DzzU^s}_1 u_{1/2}) \label{dzz}\\
\biggl(D_{zz} u\biggr)_{i,j,k-1/2}&=& \fr{1}{hz_{k-1/2}}
   \biggl(\fr{(A^z_u)_{k}}{hz_{k}} (u_{k+1/2} - u_{k-1/2}) - 
   \nonumber \\&-&
    \fr{(A^z_u)_{k-1}}{hz_{k-1}} (u_{k-1/2} - u_{k-3/2})  
   \biggr)  \quad \forall k: 2\leq k \leq K-1  
   \nonumber \\
\biggl(D_{zz} u\biggr)_{i,j,K-1/2}&=&\fr{1}{hz_{K-1/2}}\biggl[
 \alpha^{DzzU^b}_2 \fr{(A^z_u)_{K-1}}{hz_{K-1}} u_{K-1/2} -
   \nonumber \\&-&
    \alpha^{DzzU^b}_1\biggl( \fr{(A^z_u)_{K}}{hz_{K}}+  \fr{(A^z_u)_{K-1}}{hz_{K-1}}\biggr) u_{K-3/2}   \biggr]
   \nonumber   
\eeqr

The value of $hz_k$ corresponds to the difference of depths between adjacent layers  where $u,v,T,S$ are defined (see \rfg{zgrid}). The grid step $hz_{k+1/2}=\fr{hz_{k}+hz_{k+1}}{2}$ is the distance between layers where the vertical velocity $w_{k}$ and $w_{k+1}$ are defined.  The vertical grid is not uniform, so the gridsteps are not equal one to another. 

As well can see, impermeability is no longer imposed  on the bottom and the temperature and salinity of the bottom are modified. The wind tension on the top is controlled for velocities and an additional control flux of $T$ and $S$ is added on the surface. Parameters $\alpha_1$ and $\alpha_2$ helps also to control the position of the boundary modifying the depth of the first and the last layers.  

Thus, controlling boundary conditions in this paper, we control, in fact, the discretization of operators near the boundary determined by the set of coefficients $\alpha$. In this paper, we look for optimal values of $\alpha$ for the following 24 operators that approximate either derivatives or interpolations in the horizontal plane:
\begin{table}[h]
\begin{tabular}{|c|cccc|}
\hline
Argument & $u$ & $v$ & $T,s$& $\omega$\\
defined at & &&&\\
\hline
Operator&&&&\\
$D_x$ & $D_x u$ divg.  &$D_x v$ vort. & $D_x (u^2+v^2)$ adv. &$D_x\omega $ dissip.\\
      & $D_x(uS_xT,uS_xs)$&           & $D_x \rho$ press.grad.& \\    
$D_y$ & $D_y u$ vort.   &$D_x v$ divg. & $D_y(u^2+v^2)$ adv. &$D_y\omega $ dissip.\\
      &                 &$D_y(vS_yT,vS_ys)$&$D_y \rho$ press.grad.& \\ 
$S_x$ & $S_x u^2$ Kin.en.& $S_x v$ Coriolis &  $S_x s$ adv. & $S_x(S_y u)$ Coriolis\\
      &                  &                  &   $S_x T$ adv. &  $S_x \omega$ adv.  \\   
$S_y$ & $S_y u$ Coriolis &$S_y v^2$ Kin.en.  &  $S_y s$ adv. & $S_y(S_x v)$ Coriolis\\
      &                  &                  &   $S_y T$ adv. &  $S_y \omega$ adv.  \\   
\hline
\end{tabular}
\caption{Controlled  operators in $x-y$ direction}
\label{tab:horoprator}
\end{table}

Total set  of controlled $\alpha$ counts 2~000~808 coefficients. 

In the vertical direction, we control the discretization near the boundary  of 4 derivatives and 5 interpolations (see table \ref{vertoprator}) 
accompanied by two approximations of the second derivative $D_{zz}u,v$ and $D_{zz}T,s$ according to  \rf{dzz} and by the reconstruction of the vertical velocity $w$ \rf{w}. The set of controlled coefficients counts 1~197~792 elements in this case. 
\begin{table}[h]
\begin{tabular}{|c|ccccc|}
\hline
Argument &  \multicolumn{5}{|c|}{Operators}\\
\hline
$T,s$ & $D_z u$ adv.  &$D_z v$ adv. & $S_z T$ adv. &$S_z s $  &$S_z \rho$\\
$w$   & $D_z wT$ adv.  &$D_z ws$ adv. & $S_z( wD_zu)$ adv. &$S_z( wD_zv) $  &\\
\hline
\end{tabular}
\caption{Controlled  operators in $z$ direction.}
\label{vertoprator}
\end{table}

We can see the number of control variables is comparable in both cases. These numbers are also comparable with the dimension of the system state (1~707~245 variables) that we need to control identifying the initial conditions of the model.

\section{Data assimilation.}

One of the principal  purposes of variational data assimilation consists in the variation of  control parameters in order to bring the model's solution closer to the observational data. This implies the necessity to measure the distance between the trajectory of the model and the data. Introducing the cost function, we define this measure. Generally speaking, the cost function is represented by some norm of the difference between the model's solution and observations accompanied by the difference with the background that is used as regularization term.

\subsection{Cost function}
 To define the cost function we introduce dimensionless  state vector $\phi$ that is composed of five variables of the model $\phi=\{\mathbf{w_{u}} u, \mathbf{w_{v}} v,\mathbf{w_{T}} T,\mathbf{w_{s}}s, \mathbf{w_{\eta}} \eta\}^t$ weighted by coefficients $\mathbf{w}$. The choice of these weights that  are used to normalize the model variable will be discussed later. The distance between the model solution and observations is defined as the Euclidean norm of the difference
 
\beq
\xi^2=\xi^2(\phi(p,t))=\sum_{m} (({\cal H} \phi)_{m}- \phi^{obs}_{m})^2 =
\label{xiphi}
\eeq
where ${\cal H}$ is the operator that interpolates the model solution to the observation point and  the sum is performed over all available observations at time $t$. 
 
In this expression, we emphasize the implicit  dependence of $\xi$  on time and on the set of the control parameters $p$ that is composed of 
\begin{itemize}
\item the set of initial conditions of the model $ \phi_0=\{u\mid_{t=0},\; v\mid_{t=0},\;T\mid_{t=0},\; s\mid_{t=0},\; \eta\mid_{t=0}\}$,
\item  the set of the coefficients $\alpha_{xy}$ that controls the discretizations of horizontal operators (Table \ref{tab:horoprator}) in the vicinity of the continents,
\item  the set of the coefficients $\alpha_{z}$ that controls the discretizations of vertical operators (Table \ref{vertoprator}) near the bottom and near the surface,
\item Vertical diffusion coefficients $A^z_{u},A^z_{v},A^z_{T},A^z_{s}$
\end{itemize}

The cost function is composed of two terms. One of them measures the distance from the observations, and  another one is added to avoid irregular solution that may occur due to lack of observational information and the ill-posedness of the problem. Indeed, the dimension of the model state is bigger than the quantity of available observations. That means we can not identify the model state in regions where observational information is absent. To avoid this problem, we add the background term in the cost function. This term allows us to determine the solution everywhere requiring  that it must be close to the specified background.  The regularization term we use has a form
\beq
\costfun^{bgr}(p)=\sum_m (p_m-p_m^{bgr})^2 \label{costfn_bgr}
\eeq
where $p_m$ is the set of parameters under control and $p_m^{bgr}$ are background  values of these parameters. In this paper we use initial guesses as background for all parameters. That means, when we look for the optimal initial state of the model,  we use original unperturbed initial conditions as initial guess for the minimization and as the background  as well. 
Looking for optimal $\alpha$, we use combinations $\alpha_0=0$, $\alpha_1=\alpha_2=1$ for derivatives and   $\alpha_0=0$, $\alpha_1=\alpha_2=1/2$ for interpolations both for initial guess and for the background. 
 
The difference model-observations \rf{xiphi} contribute to  another  component of the cost function. Taking into account the results obtained in \cite{sw-lin}, we define the cost function as 
\beq
\costfun^{obs}(p)=\int\limits_0^T  t \xi^2(\phi(p,t)) dt \label{costfn}
\eeq
that gives a bigger  importance to the difference $\xi^2$ at the end of assimilation interval. So far, we perform the data assimilation in order to make a forecast, we need a "better" estimate of the model state at the end of the assimilation window because this state is used as the initial point for the forecast  that starts just after the assimilation. For this purpose, we require the model to go closer to observations at the end of the assimilation window increasing the weight of the distance in the cost function. 

To search for a minimum of the cost function, we shall use its gradient with respect to  control parameters. The gradient of the background  term \rf{costfn_bgr} can be calculated easily
\beq
 (\nabla\costfun)^{bgr}_n=\der{\costfun^{bgr}}{p_n}= 2(p_n-p_n^{bgr})\label{dbgrdp}
\eeq 
The  gradient of the second component of the cost function \rf{costfn} can be calculated as a Gateaux derivative of an implicit function:
 \beqr
 (\nabla\costfun^{obs})_n&=&\der{\costfun^{obs}}{p_n}=\int\limits_0^T t\biggl(  \der{\xi^2}{p_n}\biggr) dt=\int\limits_0^T t\biggl(  \sum_m\der{\xi^2}{\phi_m}\der{\phi_m}{p_n}\biggr) dt =\nonumber \\
 &=&2\int\limits_0^T t\biggl(  \sum_m (({\cal H} \phi)_m-\phi^{obs}_m)  \der{\phi_m}{p_n}\biggr) dt\label{nabj}
\eeqr 
because the  derivative $\der{\xi^2}{\phi_m}$ can easily be calculated from \rf{xiphi}: $\der{\xi^2}{\phi_m}=2 (({\cal H} \phi)_m-\phi^{obs}_m) $. The second term in \rf{nabj},  $\der{\phi_m}{p_n}$, represents the matrix of the tangent linear model that relates the perturbation of $m$th component of the model state vector $\phi_m$ to the perturbation of the parameter $p_n$. This relationship, of course, is assumed in the linear approach, that means it is only valid for infinitesimal perturbations. 

 In the classical case, when initial conditions are considered as the  only control variable, the derivative  $\der{\phi(t)}{p}=\der{\phi(t)}{\phi_0}$  is the classical tangent model that describes the temporal evolution of a small error in the initial model state. The matrix is a square matrix that is widely studied in numerous sensitivity analyses. Its singular values at infinite time limit are related to well known Lyapunov exponents that determine the model behavior (chaotic or regular) and the dimension of its attractor.
 
 In our case,  the matrix $\der{\phi(t)}{p}$ is rectangular in general. It describes the evolution of an infinitesimal error in any parameter (including initial state). However, we can study its properties   in the similar way as we do with the classical tangent linear model. Its structure and composition is described in \cite{sw-nl} for the case of using coefficients $\alpha$ as control parameters and in \cite{assimtopo} for the case when the bottom topography is used to control the model solution. 
 
The product  $\sum_m (({\cal H} \phi)_m-\phi^{obs}_m)  \der{\phi_m}{p_n}$ in \rf{nabj} represents an unusual  vector-matrix product. To calculate this product directly we would  have to evaluate all the elements of the matrix. This would require as many tangent model runs as the size of the state vector is. So, instead of the tangent model, we shall use the adjoint one that allows us to get the result by one run of the model. Backward in time adjoint model  integration that starts from $(\phi-\phi^{obs}) $  provides immediately the product  $\biggl( \der{\phi}{p}\biggr)^*(\phi-\phi^{obs}) $ which is exactly equal to  $  (\phi-\phi^{obs}) \der{\phi}{p}$ in \rf{nabj}.  

  Using these notations, we write
\beq
\nabla\costfun^{obs}= 2 \int\limits_0^T t \biggl( \der{\phi(t)}{p}\biggr)^*(\phi( p,t)-\phi^{obs}(t)) dt  \label{grad}
\eeq
where the expression in the integral is the result of the adjoint model run from $t$ to 0 starting from the vector $ (\phi( p,t)-\phi^{obs}(t))  $.

\subsection{Adjoint model}

Tangent and adjoint models have been automatically generated by the Tapenade software \cite{Hascoet04} developed by the TROPICS team in INRIA. This software analyses the source code of a nonlinear model and produces codes of its derivative $ \der{\phi}{p}$ and of the adjoint $\biggl( \der{\phi}{p}\biggr)^*$. 

The obvious advantage of the automatic tangent and adjoint code generation consists in  avoiding of  huge development work. This fact is much appreciated especially  in the case of control of internal model parameters other than initial conditions.  As it has been shown in \cite{sw-nl}, the derivative of the model with respect to boundary conditions is  composed by two  blocks: one  of them is  composed of operators acting in the space of the model variables and another one is  composed of operators linking the model variables and the controlled parameter.  The first block is responsible for the evolution of a small perturbation by the models dynamics and is similar for any data assimilation.  The second one  determines the way how this perturbation is introduced into the model and is specific to the particular parameter under control. This block is absent when the goal is to identify the initial conditions of the model because the uncertainty in initial conditions  is introduced only once, at the beginning of the model integration. But, when the uncertainty is presented in an internal model parameter, like in this case, the perturbation is introduced at each time step of  the model.  

Consequently, controlling internal model parameters we must write an additional block for the tangent model which is at least as complex as the whole tangent model for the initial point (see equations  5 and 6 in \cite{sw-nl} where these blocks are compared for a shallow-water model). Moreover, controlling internal parameters we may be brought to the necessity to enlarge the control  set, to study the influence of some other  parameter.  Simple automatic generation of tangent and  adjoint codes helps us to be free in the choice of parameters to control. 

On the other hand, automatically generated code frequently suffers from excessive requirements to computer memory. To be able to integrate the adjoint model backward in time, we have to keep the forward trajectory.  There have been proposed in \cite{Tber07} to use Griewank and Walthers binomial checkpointing algorithm \cite{griewank}, that assumes to keep  forward model solution at several  time-steps only and to recalculate all other time-steps during the adjoint model run.  The spacing  of time-steps where we keep the forward model solution should be chosen  in order to minimize the calculation time. It has been shown in \cite{Tber07} that the slowdown of the adjoint model can be limited by the factor 7 for the assimilation window of 1000 time-steps while keeping only 27 instantaneous model states.

In this paper, we start from the analysis of the adjoint code obtained from the Tapenade software and removing from keeping an unnecessary data. So far, the time-stepping is performed by the leap-frog scheme, the adjoint code saves two time levels at each step: the $n$th and the $n-1$th. However, the $n-1$th layer is  unnecessary to save because it was saved on the previous step. We avoid also saving the values of all variable on the continents because we do not need them at all. This divides the required memory approximately by two. We do not keep in memory all forcings: they have been read from files during the forward model run, they will also be read when necessary during the adjoint run. Despite the model is written with double precision, it reveals to be sufficient to keep the trajectory with the single precision only. Moreover, we can keep odd time-steps only, retrieving all even time-steps by interpolation. This interpolation  perturbs  only a little the gradient dividing the required memory by two.   

Summarizing all the efforts, we can divide  the  memory, required for keeping the forward trajectory, by 25 ensuring that 10 days assimilation window of the ORCA2 model fits into 640 MB space. Taking into account possible use of the binomial checkpointing algorithm, we can state that automatic generation of the adjoint code by Tapenade can be used even for finer resolution models and for longer assimilation windows. In these cases  we will be  limited by a long computational  time rather than by an excessive required memory. Massively parallel version of the adjoint code will, consequently,  be necessary, but it is beyond the scope of this paper. 
 
\subsection{Discontinuous diffusion coefficients} 
 
Another problem we face trying to apply variational data assimilation to the ORCA-2 configuration consists in the discontinuous coefficients of the vertical diffusion. As we have seen  in expressions \rf{az}, \rf{az-enhvisc} and \rf{az-ddm},  coefficients $A^z$ depend on the buoyancy ratio $R_\rho$ and on the Brunt-Vais\"al\"a frequency $N^2$ in a discontinuous way. Infinitesimal variation of the Brunt-Vais\"al\"a frequency from $-\varepsilon$ to  $+\varepsilon$ in some grid point results in a drastic change of all diffusion coefficients $A^z$ from values of order $10^{-5}...10^{-4}$ up to 100. Similar effect results from a small variations of the buoyancy ratio in the vicinity of 0.5 and 1. This implies the whole system is discontinuous and, hence, not differentiable. We can neither calculate the tangent model, nor its adjoint and the variational data assimilation can not be performed. This problem has already been reported in relation with the atmospheric model in \cite{geleyn}. 

So, we have to modify the diffusion coefficients in order to make the model differentiable. Unfortunately, a simple smoothing of the diffusion coefficients can not solve the problem. So far, the jumps of the coefficients values may exceed $10^6$, any reasonable smoothing modifies a lot low values of the coefficient in the adjacent points. Instead of values of order $10^{-5}...10^{-4}$ we get $10^{-2}..10^{-1}$ that dumps all the vertical physical processes. 

To be able to perform variational data assimilation, we have to suppress completely the enhanced viscosity \rf{az-enhvisc} and the double diffusive mixing \rf{az-ddm}. The  turbulent closure scheme can, however,   be smoothed replacing $A^z_{v} = \max(A^z_0,C_k  l_k  \sqrt {\bar{e}})$  in \rf{tke1} by
\beq
A^z_{v} = C_k  l_k \sqrt{\bar{e}}+\fr{A^z_0- C_k  l_k \sqrt{\bar{e}} }{1+\exp(5(C_k  l_k \sqrt{\bar{e}}-A^z_0))}
\eeq
Of course, this modifies the model solution. In particular, the suppression of the enhanced viscosity that dumps unstable situations with imaginary Brunt-Vais\"al\"a frequency may lead to uncontrolled instability in the future. The question of the smooth diffusion coefficients that we can use in the variational data assimilation  must be considered in details, but this long study is beyond the scope of this paper, where  we focus our attention on the influence of the control parameters on the solution. 

In all the matter below, we shall use the model with smooth vertical diffusion coefficients as the basic model to be compared with models subjected to modified initial conditions or modified  parametrizations of the boundary conditions. Controlling coefficients $A^z$, we shall also start from the smooth ones comparing them to the obtained in the data assimilation process. 

\subsection{Assimilated data and background fields.}

The observational data set for this model consists of the set of the ECMWF data representing sea surface anomalies  issued from Jason-1 and Envisat altimetric missions and the set of vertical profiles of the Temperature and salinity from ENACT/ENSEMBLES data banque  \cite{Ingleby}. 

The sea level anomaly  data set counts 112 thousands measured values  distributed over 20 days interval (between the 1st and the 20th of January 2006). The distribution of the data is not uniform in time: we have about 6700 points per day during first 10 days and 4500 points per day after the 10th of January. 

The temperature and salinity profiles counts about 200 000 measures during the same 20 days: more than 12000 observations per day before the the 10th of January, and about 6000 per day after this date. The quantity of data is not uniformly distributed  within each day also. Some time steps of the model (1/15 day) receive  more than 5000 data to assimilate, some other steps got 1500 data points  only. 

The background fields we use in the assimilation \rf{costfn_bgr} depend on the parameter under control.  In this paper, the initial guess for the minimization and the background were always  the same. Thus,
\begin{itemize}
\item the set of initial conditions of the model representing the model state on the January, 1, 2006, 00h GMT is used as initial guess and background when we control the initial point of the model $ \phi_0=\{u\mid_{t=0},\; v\mid_{t=0},\;T\mid_{t=0},\; s\mid_{t=0},\; \eta\mid_{t=0}\}$;
\item  the set  $\alpha_0^{xy}=0, \alpha_1^{Sxy}=\alpha_2^{Sxy}=1/2,\alpha_1^{Dxy}=\alpha_2^{Dxy}=1$ is used as initial guess and background when we  controls the discretizations of horizontal operators (Table \ref{tab:horoprator}) in the vicinity of the continents;
\item  the set of the coefficients $\alpha_0^{z}=0,  \alpha_1^{Sz}=\alpha_2^{Sz}=1/2,\alpha_1^{Dz}=\alpha_2^{Dz}=1$ that controls the discretizations of the vertical operators (Table \ref{vertoprator}) near the bottom and near the surface. This set is accompanied by $\alpha^{w}_0=0,  \alpha^{w}_1=1$ used to calculate the vertical velocity $w$ in \rf{w} and by $\alpha_0^{Dzz}=0, \alpha_1^{Dzz}=\alpha_2^{Dzz}=1$ used to approximate the second vertical derivative in \rf{dzz};
\item the instantaneous values of the vertical diffusion coefficients $A^z_{u},A^z_{v},A^z_{T},A^z_{s}$ on the 10th of January 0H GMT are used as initial guess and background when we look for optimal values of  these coefficients. 
\end{itemize}

\section{Results}

In this section we perform four experiments of data assimilation for identification of optimal values of four sets of parameters described above. All experiments are performed following the same algorithm and the same software. In each experiment one parameter is controlled, all other parameters are kept equal to their background states.   

The  data assimilation is performed by minimization of the total cost function 
\beq
\costfun(p)=10^{-4}\costfun^{bgr}(p)+\costfun^{obs}(p) \label{costfn-tot}
\eeq
composed of \rf{costfn_bgr} and \rf{costfn} using the  minimization procedure  developed by Jean Charles Gilbert and  Claude Lemarechal, INRIA \cite{lemarechal}.  This procedure uses the gradient of the cost function \rf{dbgrdp}, \rf{nabj} in the the limited memory quasi-Newton method.

We start from assimilation window $T=5$ days and allow the minimizer to make 20 calls of the routine that calculates the cost function and its gradient. Optimal values of the controlled parameters obtained as the final state of the minimization is used as initial guess in the second minimization, that is performed with the window $T=10$ days and the minimizer is allowed to make 40 calls. The set of parameters obtained as this  final result   we shall call the optimal set. 

Of course, this optimal set is optimal within the assimilation window only. However, we shall analyze the model behavior beyond the window, and namely 10 and 20 days later, on the 20th and 30th of January.

\subsection{Comparison of different control parameters}

Convergence of the cost function during the minimization procedure is shown in the \rfg{convergence}A. Despite 20 cost functions call were allowed in all experiments, the minimizer has made less iterations, i.e successful updates of the controlled parameter.   

The worst convergence is obtained when the parametrization of the lateral boundary conditions is controlled (solid line in \rfg{convergence}A.  Minimizer performs several iterations with both $T=5$ and $T=10$ days windows, but no visible reduction of the cost function is obtained. The same phenomenon we see in \rfg{convergence}B, where the evolution  of the distance $\xi(t)$   \rf{xiphi} is shown during the  assimilation and 10 days after its end. The dotted line (the model with the optimal parametrization of the lateral boundary conditions) is almost indistinguishable from the solid line that corresponds to the original model. 

This fact shows that lateral boundary conditions play no role in the ORCA-2 model. It is useless to control them hoping to improve the solution or to bring it closer to observations. This fact is in contradiction with the result obtained in \cite{sw-nl, sw-sens} where the influence of the lateral boundary conditions on the solution of  the shallow-water model is very important. But this contradiction can be evidently explained by  different  resolutions and different lateral dissipations. Indeed,  the shallow water model discussed in \cite{sw-nl, sw-sens} is an eddy resolving model with a developed turbulence while the grid step of ORCA-2 is approximately equal to 2 degrees (i.e. about 200 km) that can not be considered as eddy-resolving at all. Moreover, the lateral dissipation coefficient used for the shallow water model was either $50$ or $ 200\fr{m^2}{s}$, while in the ORCA-2 this parameter is equal to $A^{xy}_u=40000\fr{m^2}{s}$ in extra-tropical region. Such a strong dissipation admits neither  turbulence nor lateral boundary layers, i.e. configurations in which the lateral boundary conditions play an important role and worth to be controlled. 

So far, the lateral boundary conditions are not important in this case, we shall not consider them as a controlled parameter below. 

When we control either initial state $\phi_0$ or vertical diffusion coefficients $A_z$ of the model we get similar results. Control of each parameter allows  us to reduce the cost function value by more than  20\% (two dashed lines with short dashes in \rfg{convergence}A). In the \rfg{convergence}B these two parameters show close behavior too (also two dashed lines with short dashes). The distance from observations is smaller than for the model with original parameters. That means the control of the initial state of the model and its vertical diffusion coefficients is important if we want to bring the trajectory closer to observations. We should note, that the trajectory remains closer to observations even after the end of assimilation. That means we can hope to get a better forecast with the optimal model.  

\begin{figure}[h]
  \begin{center}
  \begin{minipage}[l]{0.45\textwidth}
   A. \\
  \centerline{\includegraphics[angle=0,width=0.95\textwidth]{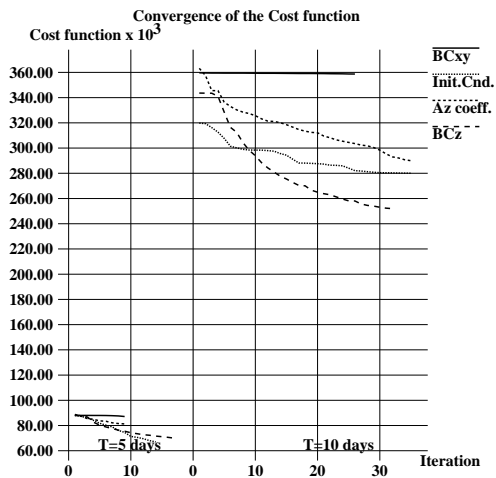}}
  \end{minipage} 
    \begin{minipage}[r]{0.45\textwidth} 
      \hfill B. 
  \centerline{\includegraphics[angle=0,width=0.95\textwidth]{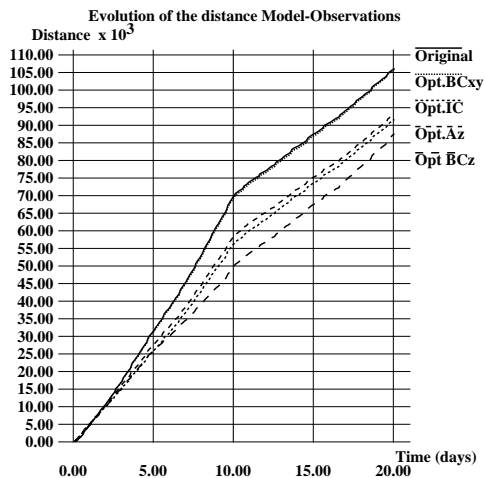}}
  \end{minipage} 
  \end{center} 
\caption{Convergence of the cost function \rf{costfn-tot} during the minimization procedure (A) and the evolution of the distance $\xi(t)$   \rf{xiphi} during  and after the assimilation (B). }
\label{convergence}  
\end{figure}

But the most influent parameter in this set of experiments is the parametrization of boundary conditions for vertical operators. If we control the conditions on the surface and on the bottom of the ocean, we get a rapid convergence of the cost function (dashed line with long dashes in \rfg{convergence}A) and the shortest distance to observations (the same line in  \rfg{convergence}B). This  indicates that vertical boundary conditions are very important in this case  from the point of view of as the  data reanalysis in the assimilation window and the forecast beyond the window.

\subsection{Modification of physical variables under control of vertical boundary conditions.}

In order to examine the modification of the physical variables produced by the use of optimal model parameters, we plot first the sea surface elevation in two regions: North Atlantic and North Pacific. These regions are characterized by the presence of strong jet-streams (Gulf Stream in the North Atlantic and Kuroshio in the North Pacific) difficult to reproduce by low-resolution models like ORCA-2. The strength and the length of the jets are usually  reduced by these models while the width is  over-estimated. 

The sea surface elevation is the variable that reflects the integral impact of the jet on the model solution. The strength of the jet can be evaluated by the orthogonal derivative of the SSH while the length is represented by the length of the region with the big gradient.

\begin{figure}[h]
  \begin{center}
  \begin{minipage}[l]{0.48\textwidth}
   A. \\ 
  \centerline{\includegraphics[angle=-90,width=0.99\textwidth]{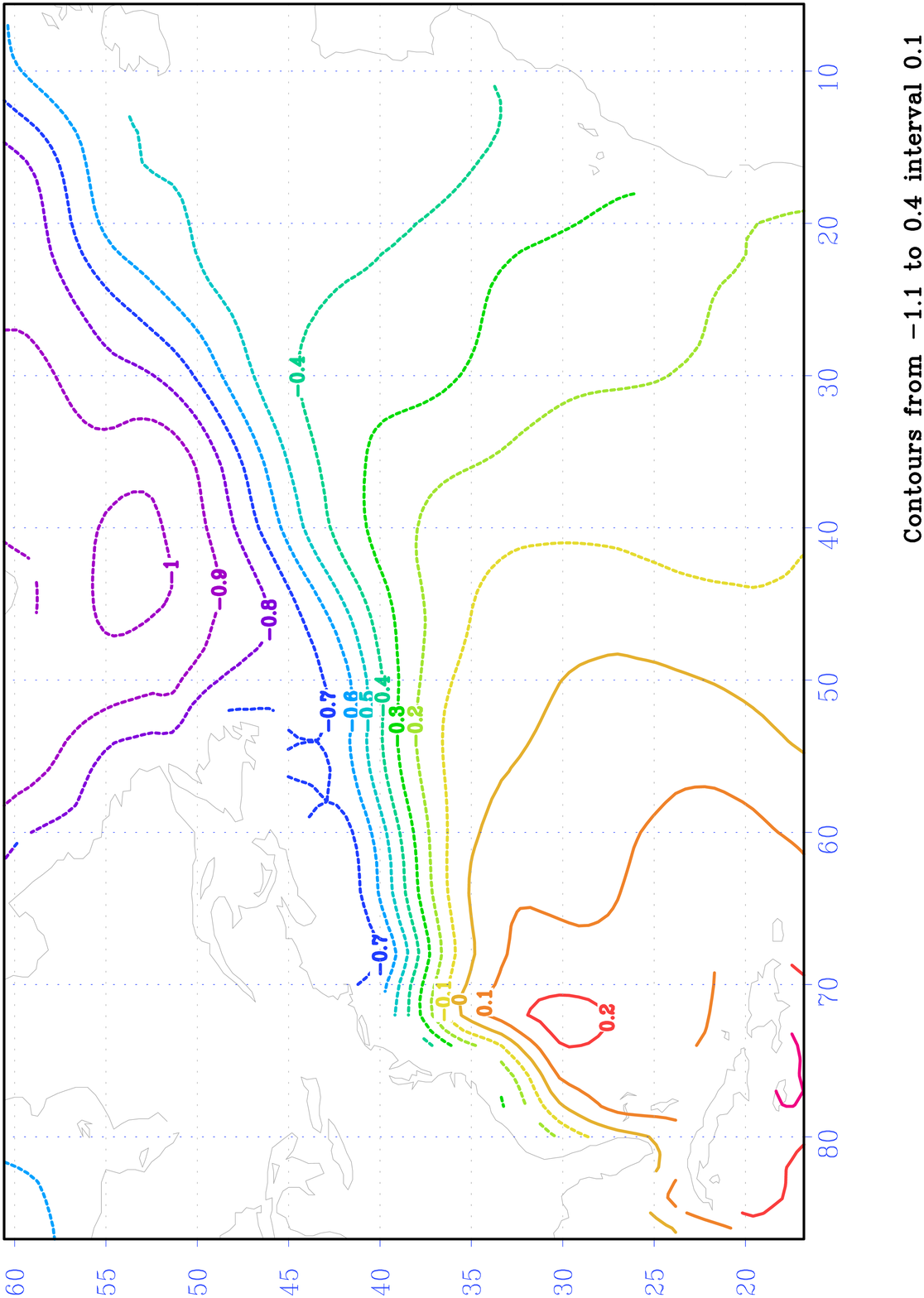}}
  \end{minipage} 
  \begin{minipage}[r]{0.48\textwidth} 
      \hfill B. 
  \centerline{\includegraphics[angle=-90,width=0.99\textwidth]{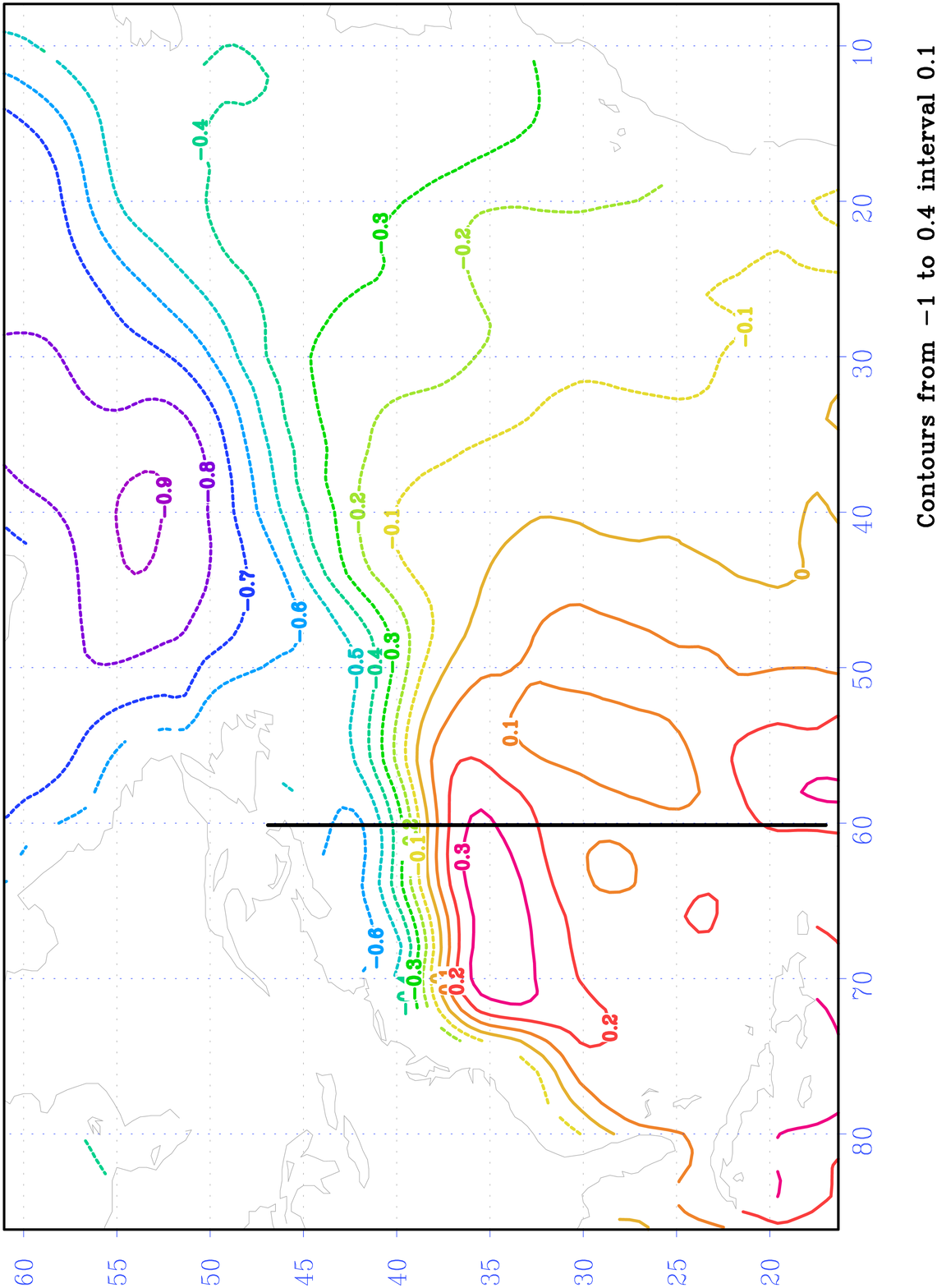}}
  \end{minipage} 
  \begin{minipage}[r]{0.48\textwidth} 
      \hfill C.
  \centerline{\includegraphics[angle=-90,width=0.99\textwidth]{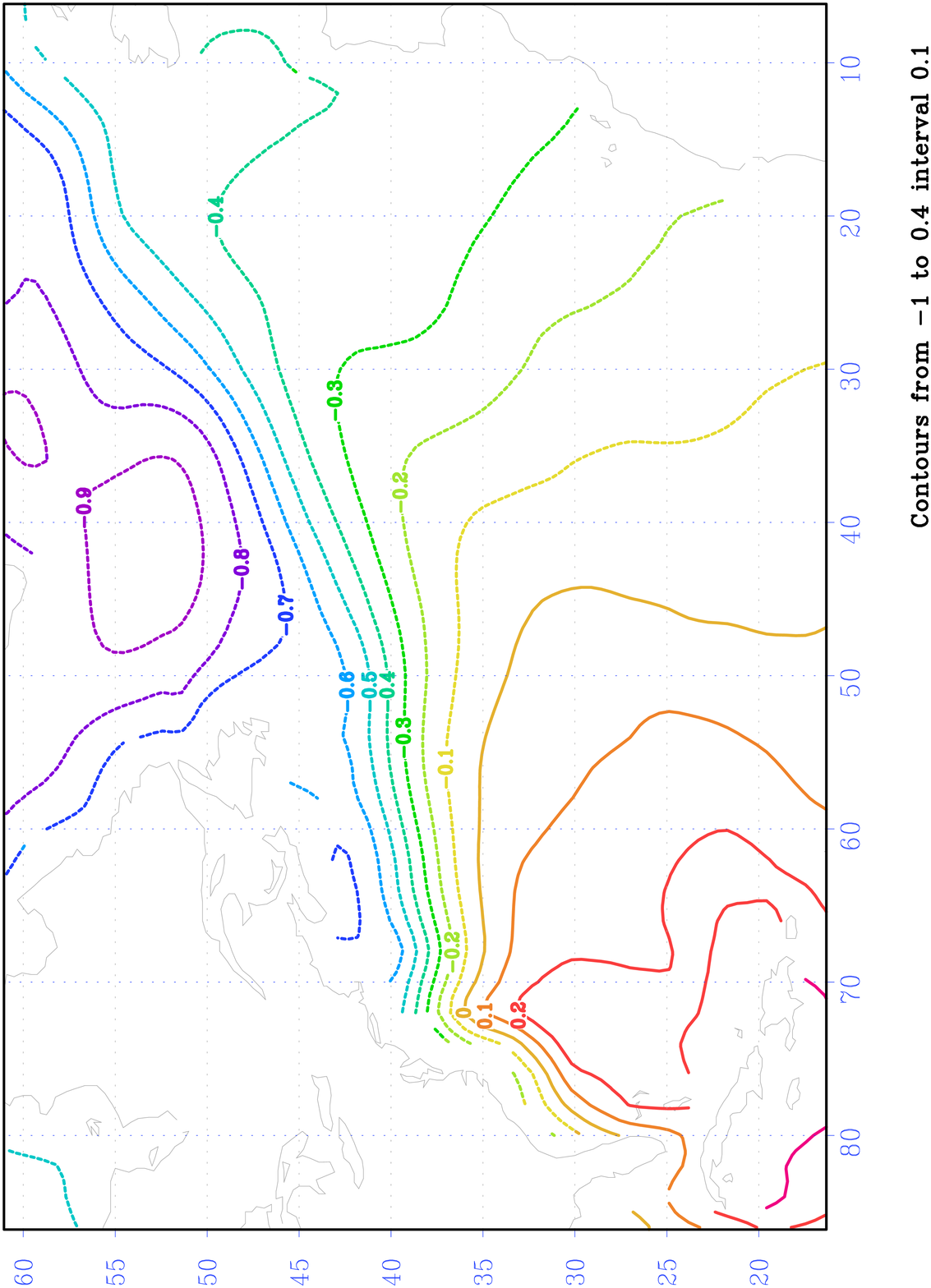}}
  \end{minipage}   
  \end{center} 
\caption{Sea surface elevation in the North Atlantic on the January, 30. Optimal initial conditions (A), Optimal Vertical boundary conditions (B) and Optimal vertical diffusion coefficients (C). }
\label{opiza-atl}
\end{figure}

Three simulation have  been performed for 30 days beginning from the 1st of January 2006. The sea surface elevation in the North Atlantic obtained as the  final states  are presented in  the \rfg{opiza-atl}. The first one (A) corresponds to  the model that  starts from  optimal initial state. The second one is obtained with the optimal parametrization on boundary conditions of all vertical operators. Optimal coefficients for vertical diffusion  were used to obtain the   third result. 

One can see that (A)  and (C) parts in the \rfg{opiza-atl} look  very similar. Optimal diffusion coefficients move the Gulf Stream slightly to the South increasing by 10 cm the positive anomaly on the South and reducing the negative anomaly on the North by the same 10 cm. The length and the strength of the Gulf Stream is  the same  in both pictures.

The modification of the circulation produced by the optimal parametrization of the vertical boundary conditions is much more visible in the \rfg{opiza-atl}B. The positive anomaly is reinforced and moved to the North-East.  The beginning of Gulf Stream is moved slightly to the North, becomes narrower and  its force  increased as well as its length. This case  clearly represents much better solution from the physical point of view. 

\begin{figure}[h]
  \begin{center}
  \begin{minipage}[l]{0.48\textwidth}
   A. \\
   \vspace{-2mm}
  \centerline{\includegraphics[angle=-90,width=0.99\textwidth]{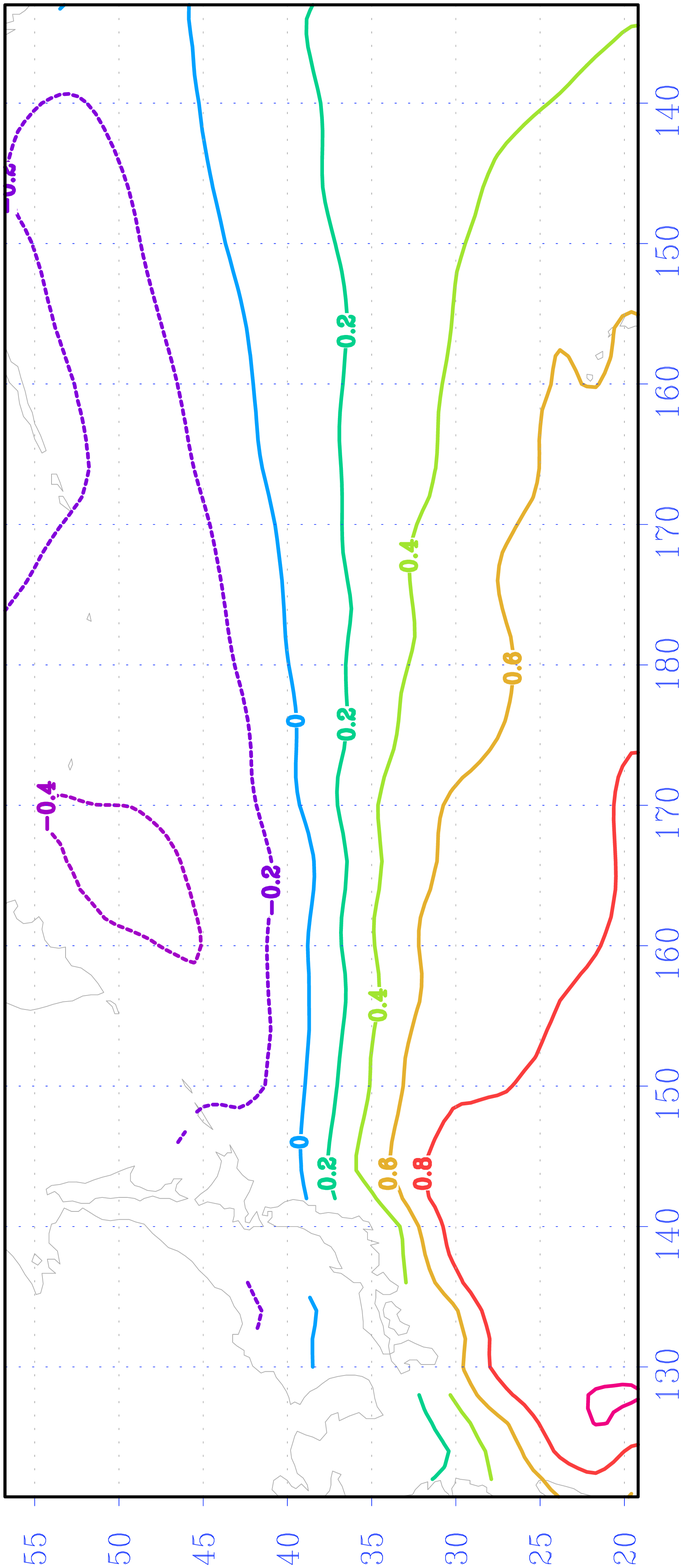}}
  \end{minipage} 
  \begin{minipage}[r]{0.48\textwidth} 
      \hfill B. 
  \centerline{\includegraphics[angle=-90,width=0.95\textwidth]{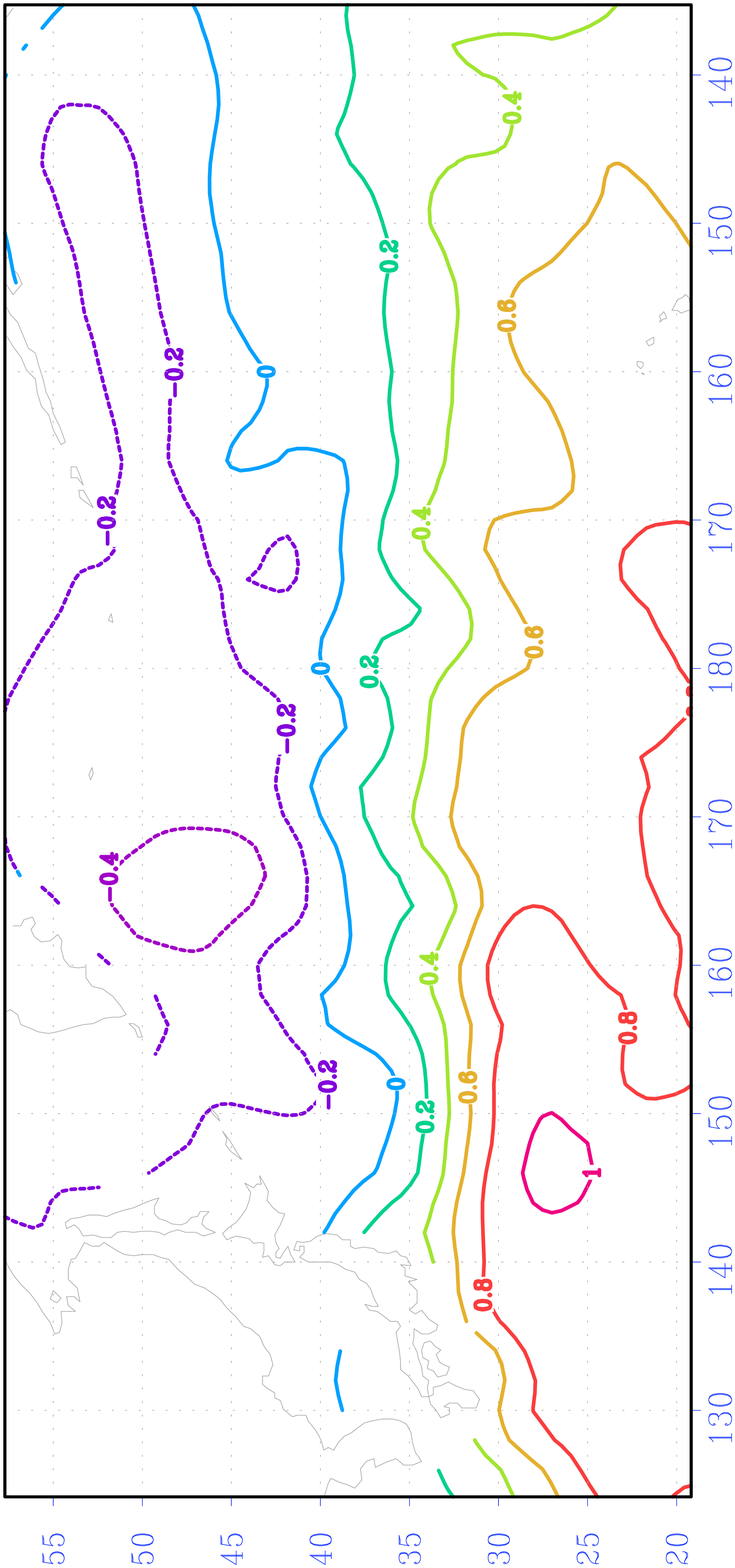}}
  \end{minipage} 
  \begin{minipage}[r]{0.48\textwidth} 
      \hfill C. 
  \centerline{\includegraphics[angle=-90,width=0.99\textwidth]{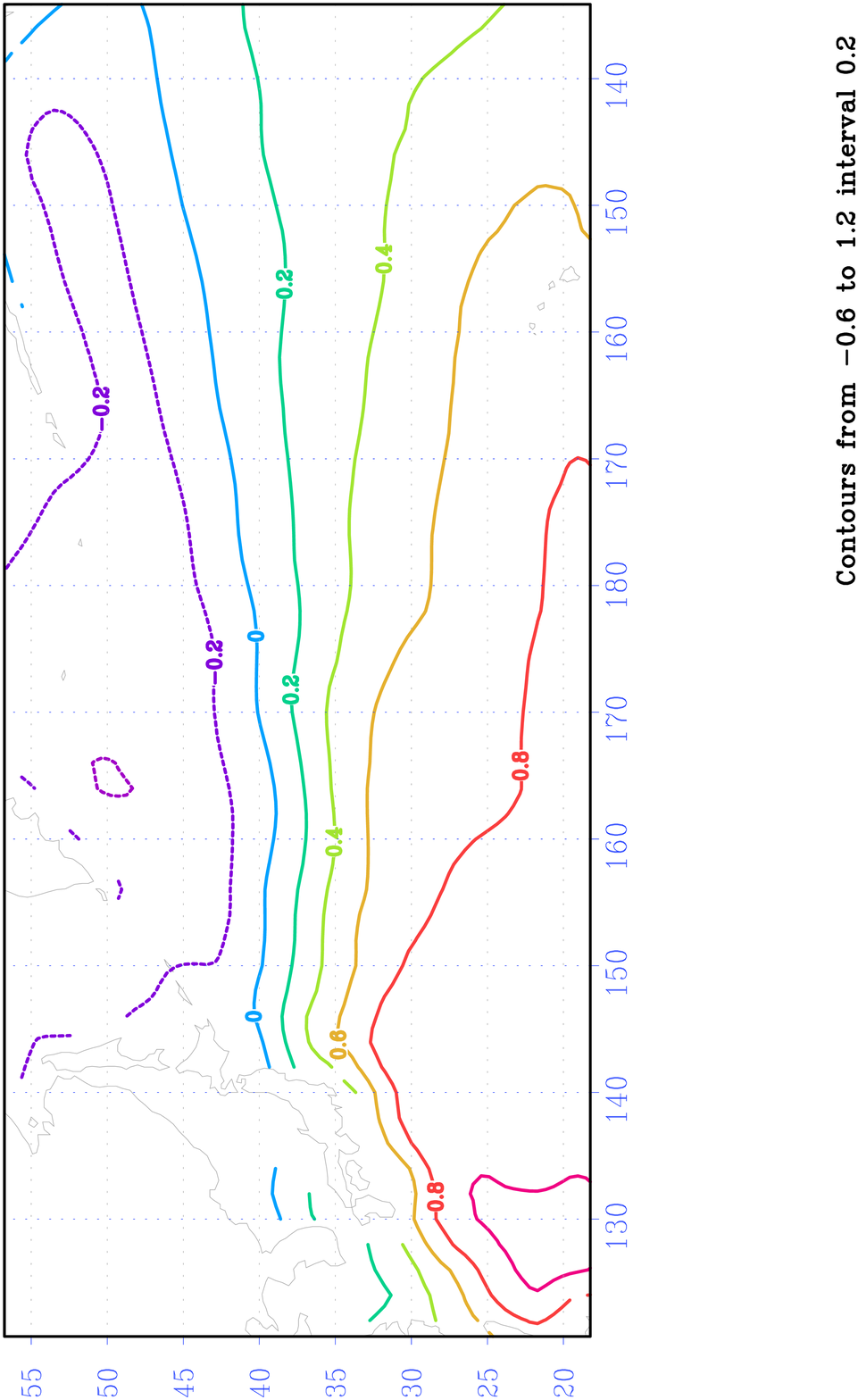}}
  \end{minipage}   
  \end{center}   
    \caption{Sea surface elevation in the North Pacific on the January, 30. Optimal initial conditions (A), Optimal Vertical boundary conditions (B) and Optimal vertical diffusion coefficients (C). }
\label{opiza-pcf}
\end{figure}

Similar effects can be observed in the North Pacific in \rfg{opiza-pcf}:  slightly reinforced positive anomaly on the South of Kuroshio  and  reduced negative one on the North in (A) and (C) parts showing similar influence of the controlled parameters on the solution. And also, the   \rfg{opiza-pcf}B obtained with the optimal boundary conditions for vertical operators is strongly modified. Like Gulf Stream, Kuroshio become longer, stronger and narrower. Moreover, it becomes also more tortuous that seems like a simulation  of moving eddies. This is also a good result obtained in frames of a coarse resolution model. 

The immediate idea one can have looking at these pictures that it is an artifact. Indeed, the sea surface elevation $\eta$ is directly related to the vertical velocity on the surface (see \rf{1.5}). We control $w$ on the surface adding control coefficients $\alpha^{w^s}_0$ and $ \alpha^{w^s}_1$ (see \rf{w}). So far, much observational data is collected by satellites and available for the sea surface elevation, the variable $\eta$ might undergo a strong forcing pulling it toward observations.  It  is possible, consequently, that  data assimilation modifies just coefficients $\alpha^{w^s}$ on the surface for $w$ velocity and this modification is immediately translated to $\eta$ with no influence on all other variables. 

However, analyzing the pattern\footnote{This picture is not presented. For more color pictures and movies, please see http://www-ljk.imag.fr/membres/Kazantsev/orca2/index.html} of  $\alpha^{w^s}$  we can not see any significant modification that can influence the jet streams. First, the magnitude of $\alpha^{w^s}_0$ is of order $10^{-7}$ while  the velocity $w$ on the surface is about $10^{-5}...10^{-4}$. Second, no characteristic patterns of  jet streams can be observed in the fields $\alpha^{w^s}_0$ and $ \alpha^{w^s}_1$. And third, the perturbation of the vertical velocity by $\alpha_1^{w^b}$ on the bottom is bigger than on the surface. 

Values of the velocity $w$ in the vertical $y-z$ section by the $60^0$W meridian are shown in \rfg{opz-w}. This section is indicated by the thick vertical line in \rfg{opiza-atl}B. One can see strong values of $w$, reaching $1.5 \fr{mm}{s}$, near the bottom of the ocean.  These values can't be observed in the original model because of impermeability boundary condition imposed for $w$ and no-slip condition for horizontal velocity components: $u\vert_{\mbox{bottom}}=v\vert_{\mbox{bottom}}=0$.

\begin{figure}[h]
  \begin{center}
  \begin{minipage}[r]{0.48\textwidth} 
  \centerline{\includegraphics[angle=-90,width=0.99\textwidth]{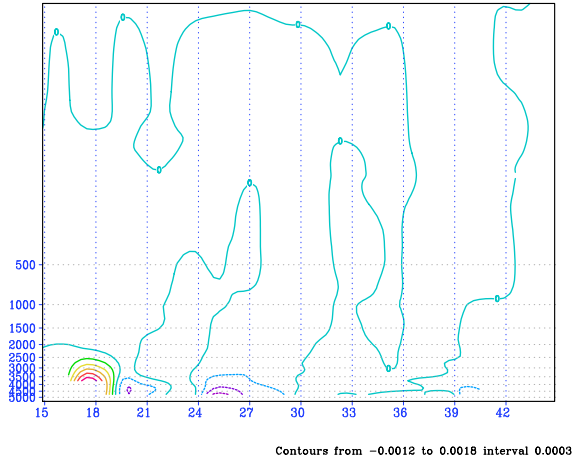}}
  \end{minipage}   
  \end{center} 
\caption{Vertical velocity $w$ obtained with optimal parametrization of vertical boundary conditions  in $y-z$ plane for $x=60^0$W. }
\label{opz-w}
\end{figure}

Consequently, we shall state that it is the modification of the bottom boundary conditions that influences the jet streams near the surface. 

An interesting question we may ask, whether it is the vertical velocity itself that is modified near the bottom, or the modification is due to  divergent modification of lateral velocities $u$ and $v$. In other words, whether coefficients $\alpha$  create   fountains at the bottom that spray out jets of water, or they create three-dimensional eddies in which the divergence of lateral velocities $\xi$ \rf{1.6} is balanced by $\der{w}{z}$.

\begin{figure}[!h]
  \begin{center}
  \begin{minipage}[l]{0.48\textwidth}
   A. \\
  \centerline{\includegraphics[angle=-90,width=0.99\textwidth]{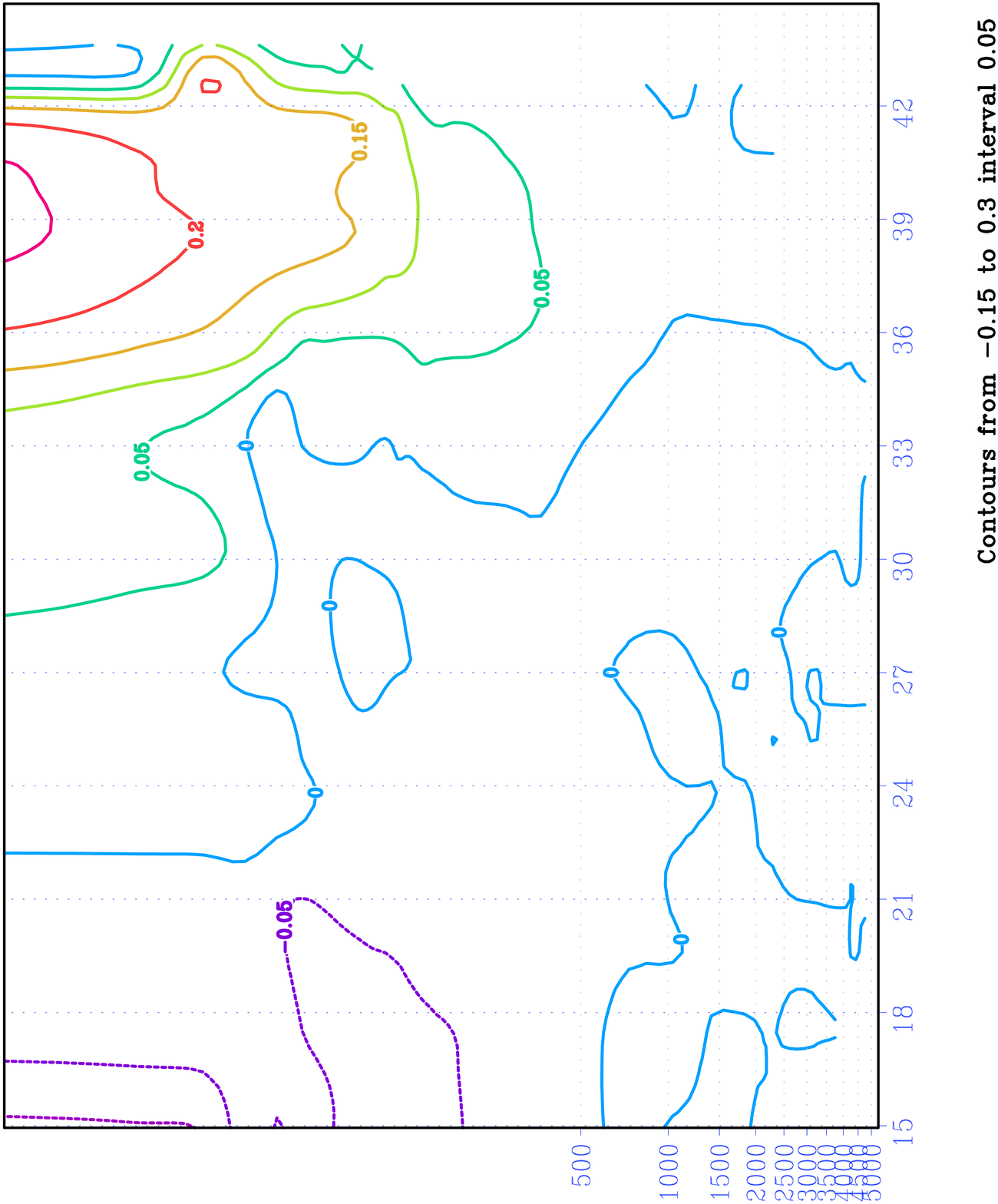}}
  \end{minipage} 
  \begin{minipage}[r]{0.48\textwidth} 
      \hfill B. 
  \centerline{\includegraphics[angle=-90,width=0.99\textwidth]{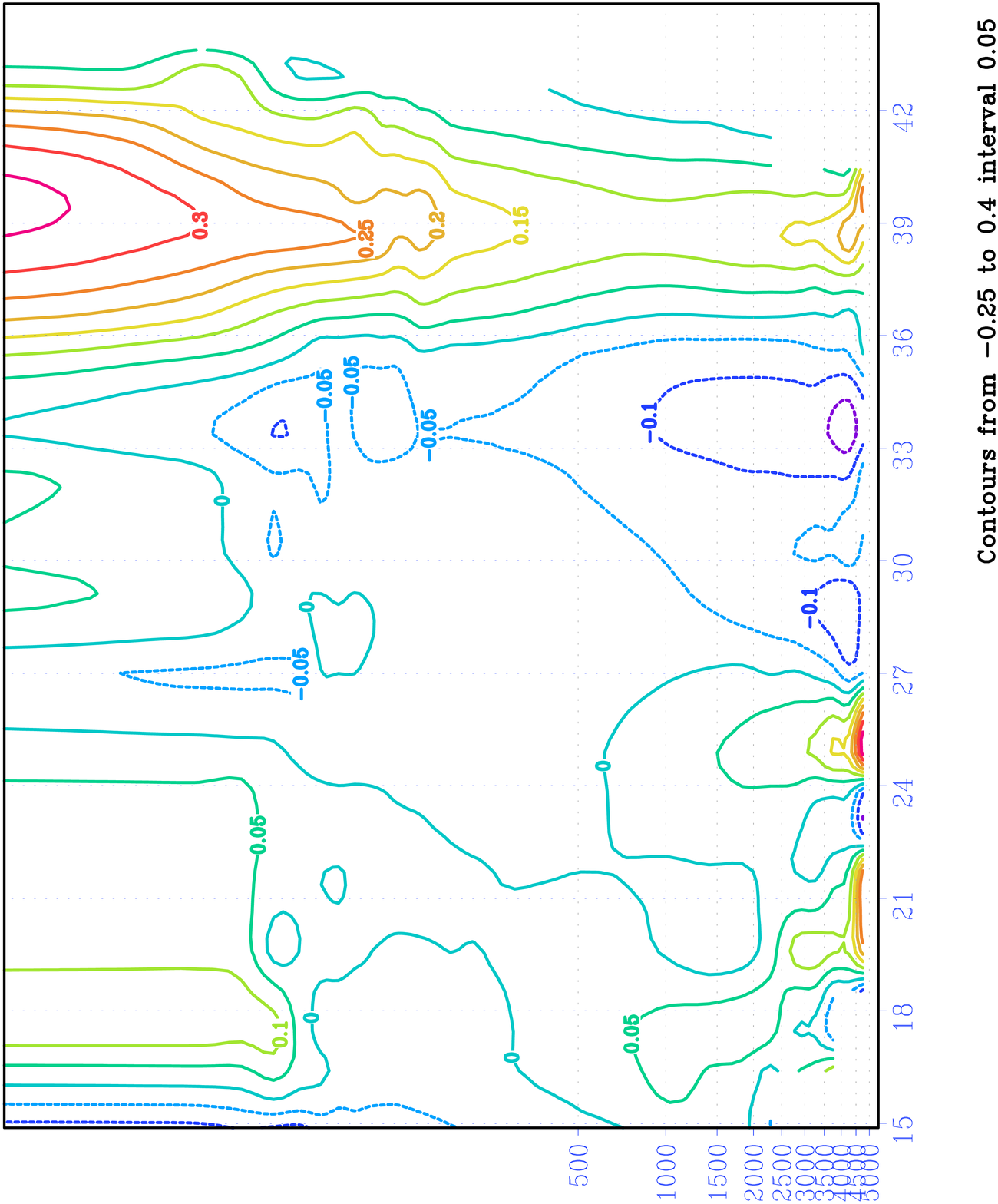}}
  \end{minipage} 
  \end{center} 
\caption{Velocity component $u$  in $y-z$ plane for $x=60^0$W: obtained with optimal initial conditions (A) and optimal parametrization of vertical boundary conditions (B).}
\label{opz-u}
\end{figure}

To answer this question, we plot the $u$ velocity in the same $y-z$ section in \rfg{opz-u}B (velocity $u$ is orthogonal to the picture plane) and we  see that the velocity $u$ also has anomalies near the bottom in the same places as the velocity $w$, and namely in regions  $18^0...21^0$N and $24^0...27^0$N. That indicates the source of the $w$ velocity may lie in the divergent part of the lateral velocity components. Moreover, in \rfg{opz-u}B, one can see an   important anomaly of $u$ near the bottom ($39^0...41^0$N) that is almost non divergent (the derivative  $\der{u}{x}$ is compensated by $\der{v}{y}$) producing only a little anomaly in $w$. 

Moreover, in \rfg{opz-u}B one can easily see the influence of  anomalies of $u$ and $v$ near the bottom on the jet stream near the surface. The  positive anomaly of $u$ just below the jet stream ($39^0...41^0$N) maintains the flux over the whole water column ensuring more than $10\fr{cm}{s}$ velocity at the 500 m depth. Comparing the velocity pattern with the velocity obtained in the experiment with optimal initial conditions presented in \rfg{opz-u}A, we see that $u$ can hardly reach the value of $5\fr{cm}{s}$  at the 500 m depth. That means the Gulf Stream become not only stronger (one can see the surface velocity exceeds $35\fr{cm}{s}$ instead of $25\fr{cm}{s}$ in \rfg{opz-u}A) and longer, but also deeper. 

This fact brings us to the conclusion that the major impact of the data assimilation is made on the parametrization bottom boundary conditions for lateral velocities. Consequently, the modification of the sea surface elevation in the jet-stream regions is not a simple artefact, but the result of optimization of the deep dynamics.

Finally, we should note that the temperature and salinity fields has been modified only a little in the data assimilation. We can see almost no difference with the original fields near the bottom, and a little difference near the surface, especially  in the equatorial region.

As we have seen in \rfg{convergence}, optimal coefficients $A^z$ allows us to bring the model solution closer to observations. However, they don't amplify the stream jets \rfg{opiza-atl}, \rfg{opiza-pcf} like optimal boundary conditions. Analyzing the modifications made by the data assimilation, we can see numerous small regions (and even points) where the viscosity become vanishing.  This result seems not to be physically interesting. 

Of course, more pictures should be presented in this section in order to show in details the modification of different variables of the model. However, preserving the volume of the paper and taking into account the impossibility to present multimedia content in the Journal, the reader can find numerous color pictures and movies at http://www-ljk.imag.fr/membres/Kazantsev/orca2/index.html 

\section{Conclusion and perspectives}

This paper  shows possible advantages of extending of the now classical variational data assimilation procedure to the control of the  model parameters other than initial conditions. Controlling parametrization of boundary conditions of  vertical operators in the ORCA-2 model we have brought the solution closer to observations not only during the assimilation stage, but also during the forecast. We have shown also that the jet streams in the North Atlantic and the North Pacific are modified under control of the vertical boundary conditions. This result seems to be interesting because these jets are extremely difficult to simulate in frames of a low-resolution model. Another  important particularity consists in the fact that it is the bottom boundary conditions for lateral velocities $u$ and $v$ were subjected to the major modifications.  That shows the importance of the appropriate approximation of the bottom topography and processed in the bottom boundary layer. 

Performing the control of parameters, it is not clear in advance what parameter is the most important for a given model. It was lateral boundary conditions in the case of the shallow-water model in \cite{sw-nl, sw-sens}, but they play no role in the ORCA-2 model.  Taking into account the effort made to construct the vertical diffusion coefficients $A^z$ for this model  (including an additional  evolution equation for TKE $\bar{e}(x,y,z,t)$), one can suppose  they might be extremely important in this configuration. However, their influence is not as strong as the influence of vertical boundary conditions. 

This fact helps us to understand that boundary conditions are only important to optimize in the case when the corresponding operator is important for the model dynamics. The solution of the shallow-water model discussed in \cite{sw-nl, sw-sens} was exhibiting either developed turbulence or strong under-resolved boundary currents showing that horizontal operators and their boundary conditions should be considered carefully.  In the case of ORCA-2, almost all the  horizontal dynamics is dumped by the strong dissipation that leads to a little influence of corresponding boundary conditions. 

But, in the case of ORCA-2, the vertical dynamics is important. The dissipation and diffusion coefficients are chosen as low as possible (sometimes as low as $A^z=10^{-4} \fr{m^2}{s}$ compared to $A^{xy}=4\tm 10^{+4} \fr{m^2}{s}$ ) in order to preserve the vertical currents. This fine adjustment points out the importance of the vertical dynamics and requires, consequently, careful treatment of its boundary condition. The present study confirms this fact, showing the optimizations of the vertical boundary conditions modifies much the currents.

Summarizing, we can say that the control of model parameters can help us 
\begin{itemize}
\item to compensate the model errors and to bring  the model solution closer to observational data improving the forecast;
\item to see what parameters are more or less  important for a given model and for a given configuration; 
\item to see what geographical region requires a particular attention in formulation of the boundary conditions.
\end{itemize}

The use of the automatic differentiation tool reveals to be extremely useful in this study helping us to avoid the huge   coding and debugging work. This fact is very appreciated in the situation when we come out the frames of controlling the only classical model parameter --- its initial state. Now, we can be brought to consider various parameters of the model as  worthy to be optimized, facing the necessity to get the derivative of the model and its adjoint with respect to the chosen parameter. 

The major shortcoming of automatic differentiation tool that consists in an  excessive requirement  of computer memory (hundreds or thousands times of the model code)  can be avoided by the memory usage optimization and by    the binomial checkpointing algorithm \cite{griewank}. These techniques bring the memory demand into a reasonable limit and the principal barrier to overcome becomes the usual computing time. Consequently, automatic  differentiation of massively parallel codes and parallelization of the adjoints is necessary. 

However, the present study can not be considered as a finalized control of the model parameters. Numerous  points, that are extremely important from physical and mathematical points of view, have not been discussed in this paper. 

Standing on  the mathematical point of view, we must address two principal questions about uniqueness and stability of the parameter identification. These questions arise because we are  solving  an inverse problem, that is non-linear and probably ill-posed.  The uniqueness of the solution determines whether it is at all possible to obtain the value of unknown parameter from  observational data. The notion of stability determines whether  small errors in data would cause serious errors in the identified  parameter. 

 Both these questions have negative answer in this paper. The set of control coefficients was intentionally chosen to be too large to ensure an unique solution.  As it has been shown in \cite{assimbc1}, exceedingly large set of $\alpha$ leads to a    non-null kernel   of the Hessian and   results in a non-unique  choice of optimal boundary conditions. 
 
 Moreover, the nonlinearity of the model solution with respect to boundary conditions is far from being quadratic. Consequently, the cost function $\costfun(p)$ \rf{costfn-tot}, along with the global minimum, may possess numerous local ones. Applying the minimization procedure, we shall find one of them instead of the global one that will result in a supplementary non-uniqueness of the solution. Incremental data assimilation technique may help us to improve the minimization by finding a deeper minimum, but it can not guarantee the minimum found is global.   
 
 And finally, we can not find a minimum at all because we can perform only a  limited  number (few) iterations of the minimization process in a reasonable computing time.  Even with ORCA-2, one of the simplest models of the Global Ocean, we could perform some 40 iterations. Hence, the obtained result is far from even a local minimum. 
 
 Consequently, this study can not pretend to be an identification of the  boundary conditions, but rather a way to compensate some model errors due to the lack of resolution and approximative parametrizations. 
 
The same conclusion we can  make  from the physical point of view also. First of all, coefficients $\alpha$ are not physical parameters. They have been chosen as controls because any imaginable  boundary condition  can be approximated by a combination of these three $\alpha$. Even unphysical conditions, like a permission to cross the continent contours, are intentionally accepted because they may point out that the model geometry is not in the agreement with the model dynamics. Moreover, we accept different $\alpha$ for different operators even dealing with the same variable assuming the possibility of different boundary conditions imposed to the same variable and increasing the probability to get   a non-null kernel   of the Hessian  when one  $\alpha$ compensates one another. 

  Thus, an exceedingly wide set of $\alpha$ is controlled instead of some physical set of boundary conditions in order to allow  the data assimilation to modify any  $\alpha$ bringing the model solution closer to observations. 

Consequently, the values of  $\alpha$ them-self  are neither  important, nor physically meaningful in this study. That's why they are not plotted in this paper. But they show the result we can potentially get optimizing boundary conditions, the regions where this optimization is particularly important and which operators and which variables of the model should attract a special  attention.  Analyzing the modification of the physical variables, we can further reduce the quantity of the controlled $\alpha$ and proceed to the control of the physical boundary conditions including the position of the boundary that may should be moved from the current position.  

Acknowledgement: The author would like to express his gratitude to  Arthur Vidard for providing the ORCA-2 configuration of the NEMO accompanied by  forcings and  observational data. All the contour pictures have been prepared by the Grid Analysis and Display System (GrADS) developed in  the Centre for Ocean-Land-Atmosphere Interactions,   Department of Meteorology, University of Maryland.

\bibliography{/home/kazan/text/mybibl}

\end{document}